\documentclass{emulateapj}

\usepackage{graphicx}
\usepackage{amsmath}
\usepackage{natbib}
\usepackage{amssymb}

\usepackage[breaklinks,colorlinks,urlcolor=blue,citecolor=blue,linkcolor=blue]{hyperref}
\usepackage{hyperref}

\begin{document}

\newcommand{\be}{\begin{equation}}
\newcommand{\ee}{\end{equation}}

\title{Effects of Fall-Back Accretion on Proto-Magnetar Outflows in Gamma-Ray Bursts and Superluminous Supernovae}

\author{Brian D.~Metzger$^{1}$, Paz Beniamini$^{2}$, Dimitrios Giannios$^{3}$}
\altaffiltext{1}{Department of Physics and Columbia Astrophysics Laboratory, Columbia University, New York, NY 10027, USA.  email: bdm2129@columbia.edu}
\altaffiltext{2}{Department of Physics, The George Washington University, Washington, DC 20052, USA}
\altaffiltext{3}{Department of Physics and Astronomy, Purdue University, 525 Northwestern Avenue, West Lafayette, IN 47907, USA}

%\maketitle

\begin{abstract} Rapidly spinning, strongly magnetized proto-neutron stars (``millisecond proto-magnetars") are candidate central engines of long-duration gamma-ray bursts (GRB), superluminous supernovae (SLSNe), and binary neutron star mergers.  Magnetar birth may be accompanied by the fall-back of stellar debris, lasting for seconds or longer following the explosion.  Accretion alters the magnetar evolution by (1) providing an additional source of rotational energy (or a potential sink, if the propeller mechanism operates); (2) enhancing the spin-down luminosity above the dipole rate by compressing the magnetosphere and expanding the polar cap region of open magnetic field lines; (3) supplying an additional accretion-powered neutrino luminosity that sustains the wind baryon-loading, even after the magnetar's internal neutrino luminosity has subsided.  The more complex evolution of the jet power and magnetization of an accreting magnetar more readily accounts for the high $^{56}$Ni yields GRB SNe and irregular time evolution of some GRB light curves (e.g.~bursts with precursors  followed by a long quiescent interval before the main emission episode).  Additional baryon-loading from accretion-powered neurino irradiation of the polar cap lengthens the timeframe over which the jet magnetization is in the requisite range $\sigma \lesssim 10^{3}$ for efficient gamma-ray emission, thereby accommodating GRBs with ultra-long durations.  Though accretion does not significantly raise the maximum energy budget from the limit of $\lesssim few \times 10^{52}$ ergs for an isolated magnetar, it greatly expands the range of magnetic field strengths and birth spin periods capable of powering GRB jets, reducing the differences between the magnetar properties normally invoked to explain GRBs versus SLSNe.
\end{abstract}

\section{Introduction}

Rapidly-spinning, strongly-magnetized neutron stars (NS) formed in core collapse supernovae (``millisecond proto-magnetars")  are promising candidates for the central engines of long-duration gamma-ray bursts (GRB; \citealt{Usov92,Dai&Lu98,Wheeler+00,Zhang&Meszaros01,Thompson+04,Bucciantini+07,Uzdensky&MacFadyen07,Metzger+07,Metzger+11,Mazzali+14,Vlasov+14,Vlasov+17,Beniamini+17}) and hydrogen-poor superluminous supernovae (SLSNe; \citealt{Kasen&Bildsten10,Woosley10,Dessart+12,Chatzopoulos+13,Metzger+14,Sukhbold&Thompson17}).  Powering a relativistic GRB jet by the magnetic dipole spin-down luminosity of an isolated (non-accreting) magnetar requires a large surface magnetic field strength of $B_{\rm d} \gtrsim few \times 10^{15}$ G, while SLSNe instead typically require weaker fields of $B_{\rm d} \lesssim few \times 10^{14}$ G (e.g.~\citealt{Metzger+15}).  This difference arises because the spin-down timescale over which the magnetar's rotational energy $\gtrsim 10^{52}$ ergs is extracted must be short in  GRBs to explain their typically observed durations of $\lesssim$ tens of seconds.  By contrast, substantially enhancing the optical luminosity of the supernova, as needed to explain SLSNe, instead requires the magnetar's energy be deposited into a nebula behind the ejecta over longer timescales $\gtrsim$ days, comparable to the photon diffusion timescale out of the ejecta.  

A millisecond magnetar is also created from the coalescence of binary NSs (e.g.~\citealt{Price&Rosswog06,Siegel+13,Kiuchi+14}).  A merger remnant that avoids promptly collapsing into a black hole can power long-lived electromagnetic counterparts to the gravitational wave emission, either by supplying an additional source of mildly-relativistic radioactive kilonova ejecta (on timescales $\lesssim few$ s after the merger; \citealt{Metzger+18}), relativistic ejecta and prompt gamma-ray/X-ray emission (on timescales of $\sim 1-10^{3}$ s; e.g.~\citealt{Metzger+08,Bucciantini+12,Rowlinson+13,Zhang13}), or by enhancing the optical/X-ray emission of the kilonova via spin-down luminosity (on timescales $\sim$ hours$-$days; \citealt{Kulkarni05,Yu+13,Metzger&Piro14,Gao+15,Siegel&Ciolfi16a,Siegel&Ciolfi16b}), analagous to a scaled-down version of SLSNe.  The fact that no such emission was observed following the recent LIGO NS merger GW170817 \citep{LIGO+17DISCOVERY,LIGO+17Fermi} favored a relatively rapid black hole formation in this event \citep{Margalit&Metzger17,Pooley+18,Margutti+18}.

While the birth of millisecond magnetars can successfully reproduce many of the observed properties of GRBs (e.g.~\citealt{Metzger+11,Beniamini+17}) and SLSNe (e.g.~\citealt{Nicholl+17,Yu+17}), several outstanding challenges remain:
\begin{itemize}
\item{{\bf Energetics.}  The total energy of the GRB and its afterglow in the magnetar model cannot exceed a fraction of the maximum rotational energy of a NS, which is $\sim 3\times 10^{52}$ ergs for a canonical NS of mass 1.4$M_{\odot}$ (\citealt{Metzger+15,Beniamini+17}).  This limit is comparable to the measured or inferred energies of a handful of long GRBs and their afterglows (e.g.~\citealt{Cenko+11,Beniamini2015}).  It is also close to being violated by the very luminous optical transient ASASSN-15lh (\citealt{Dong+16}), which radiated $\gtrsim 10^{52}$ ergs and was identified as a SLSN (see, however \citealt{Leloudas+16,Margutti+17}). }
\item{{\bf Radioactive Nickel.}  The broad-lined Type Ic supernovae which accompany long GRBs (e.g.~\citealt{Woosley&Bloom06,Liu+16}) are inferred to produce a large quantity $\gtrsim 0.3M_{\odot}$ of $^{56}$Ni (e.g.~\citealt{Mazzali+14}).  If this radioactive isotope is synthesized by shock-heating of the inner ejecta layers of the progenitor star to the requisite high temperatures $\gtrsim 5\times 10^{9}$ K, then the kinetic luminosity of the central engine's outflow must be $\gtrsim 10^{51}-10^{52}$ ergs s$^{-1}$ over the first $\lesssim 1$ s following the explosion (e.g.~\citealt{Suwa&Tominaga15,Wang+16,Barnes+17}).  Though an isolated magnetar with a very strong surface magnetic field $\gtrsim 10^{16}$ G can produce such a high jet luminosity, its rapid spin-down would leave the engine with little rotational energy remaining to power a luminous GRB jet over the following minute.\footnote{Though we note that the spin-down luminosity of even an isolated magnetar will not be constant at early times $\ll t_{\rm sd}$, where $t_{\rm sd}$ is the spin-down time, if the open magnetic flux threading the magnetar surface shrinks as the wind magnetization grows in time (\citealt{Bucciantini+06}; see eq.~\ref{eq:edotsd}).}}
\item{{\bf Magnetar Parameters in GRB versus SLSNe.}  The magnetic fields $B_{\rm d} \gtrsim few\times 10^{15}$ G of  magnetars required to explain the luminosities of long GRBs greatly exceed those $B_{\rm d} \lesssim few \times 10^{14}$ G required to power SLSNe.  As both models require similar birth spin periods, and the likely source of the strong magnetic field is the free energy available in rotation (e.g.~\citealt{Duncan&Thompson92}), it is not clear what additional property gives rise to such a disparate range of magnetic field strengths (corresponding to a difference of a factor of $\gtrsim 100$ in magnetic energy of the dipole component).  Many GRBs are also accompanied by extended X-ray plateaus \citep{Burrows+06}, which have been interpreted as being powered by the spin-down of long-lived magnetars (e.g.~\citealt{Zhang&Meszaros01}); however, the magnetic field strength required to explain the plateau luminosity and duration are generally lower than those required to power the GRB jet at earlier times (\citealt{Rowlinson+13,Gompertz+14,Rowlinson+14,Lu&Zhang14}).  Time-evolution of the magnetic field strength can be invoked, but fine-tuning may be required to account for the large energy of both the prompt and plateau phases \citep{BeniaminiMochkovitch2017}. }
\item{{\bf Ultra-Long GRBs.} In addition to canonical long GRBs, a handful of bursts with much longer durations $\gtrsim 10^{3}$ s have been discovered \citep{Gendre+13,Levan+14,Boer+15,Levan15}.  It remains debated whether these ``ultra-long GRBs" are simply the long-duration tail of a single long GRB population, or whether they represent a distinct class with potentially different progenitors.  One clue was provided by GRB 111209A, a ultra-long burst of duration $T_{\rm GRB} \approx 10^{4}$ s, followed by a supernova with a luminosity between those of normal GRB SNe and SLSNe \citep{Greiner+15,Kann+16}.  \citet{Metzger+15} showed that the same magnetar engine, with a magnetic field strength $B_{\rm d} \approx 3\times 10^{14}$ G and dipole spin-down timescale of $t_{\rm sd} \sim 10^{4}$ s $\sim T_{\rm GRB}$, could in principle power both a GRB jet (at the observed luminosity and duration) while also enhancing the luminosity of the subsequent supernova to the observed level (see also \citealt{Gompertz&Fruchter17,Margalit+18}).  \\

However, the production of a jet with the correct power output and timescale is of itself not sufficient to explain the prompt gamma-ray emission; the jet magnetization $\sigma$, which is inversely proportional to the baryon loading (see eq.~\ref{eq:sigma} for definition), must also lie in a critical range $100 \lesssim \sigma \lesssim 3000$ for the jet's Poynting flux to be efficiently converted into MeV radiation (e.g.~\citealt{Beniamini&Giannios17}).  Following the supernova explosion, a proto-NS radiates its gravitational binding energy through neutrino emission during its Kelvin-Helmholtz contraction \citep{Burrows&Lattimer86}, becoming optically-thin to neutrinos on a timescale $t_{\rm thin} \approx 10-30$ s \citep{Pons+99}.  These neutrinos drive a substantial flux of baryons  from the NS surface \citep{Qian&Woosley96}, keeping the jet magnetization in the optimal range to power GRB emission at times $\lesssim t_{\rm thin}$; however, following the optically-thin transition $t \gtrsim t_{\rm thin}$ the proto-NS neutrino luminosity plummets and the jet magnetization abruptly rises to enormous values (closer to those of normal pulsar winds), causing the prompt gamma-ray emission to cease \citep{Metzger+11,Beniamini+17}.  While the agreement between $t_{\rm thin}$ and the durations of most long GRBs lends support to the magnetar model \citep{Metzger+11}, for the same reason it then becomes problematic to account for ultra-long GRBs with prompt emission lasting for times $T_{\rm GRB} \gg t_{\rm thin}$.}
\item{{\bf Irregular GRB Light Curves (e.g.~Precursor Gaps).}  Although all GRB light curves show variability, some of this behavior can be attributed to the central engine or variations in the jet properties imprinted by its propagation through the progenitor star (e.g.~\citealt{Morsony+10}).  However, some GRB light curves are characterized by initial ``precursor" emission, which is followed by large gaps in time of up to hundreds of seconds before the onset of the main emission episode (e.g.~\citealt{Lazzati05,Wang&Meszaros07,Wu+13}).  Since the spin-down evolution of an isolated magnetar is relatively smooth, and is not expected to turn off for an extended period of time, such ``gapped" light curve behavior is challenging to accommodate in the magnetar picture.
}  

\end{itemize}

This paper explores to what extent the above issues can be alleviated by considering the effects of fall-back accretion on the evolution of millisecond magnetars and the properties of their relativistic jets.  Core collapse supernovae are generally expected to leave a fraction of the stellar progenitor bound to the central compact object following the supernova explosion, which will fall-back over an extended period of time and  circularize into an accretion disk (\citealt{Chevalier89,Zhang+08,Fryer+09,Woosley&Heger12,Quataert&Kasen12,Perna+14}).
Binary NS mergers will also in general be accompanied by accretion from a remnant torus of high-angular momentum material (e.g.~\citealt{Fernandez&Metzger13}) and by delayed fall-back from the cloud of matter ejected by shocks and tidal forces dynamically during the merger (e.g.~\citealt{Rosswog07,Fernandez+17}).

One role of accretion is to supply additional angular momentum to the magnetar, in principle enhancing the reservoir of rotational energy available to power the GRB jet or SLSNe (\citealt{Piro&Ott11,Bernardini+13,Bernardini+14,Gompertz+14}).  On the other hand, torques from the disk on the star may also result in the loss of stellar angular momentum if accretion occurs in the ``propeller" regime.   Likewise, the addition of too much mass will cause the magnetar to collapse to a black hole, abruptly truncating the source of the magnetar wind feeding the GRB jet \citep{Rowlinson+10} or SLSNe (\citealt{Moriya+16}).  By pushing the inner edge of the disk inside the light cylinder, a sufficiently high accretion rate acts to enhance the spin-down rate of the magnetar by opening up a large fraction of its magnetosphere (e.g.~\citealt{Parfrey+16}) as well as increasing the baryon-loading of the jet through enhanced neutrino-heated outflows.  These effects can change the temporal evolution of the magnetization and power of the GRB jet; the latter for instance no longer necessarily obeys the standard $\propto t^{-2}$ spin-down decay predicted at late times for an isolated magnetar.

This paper is organized as follows.  In $\S\ref{sec:model}$, we describe our model for the evolution of proto-magnetars and their relativistic jets, accounting for the effects of mass fall-back.  In $\S\ref{sec:results}$, we present our results for the magnetar spin-down evolution (\S\ref{sec:spindown}) and implications for some of the issues raised above, including $^{56}$Ni production in GRB SNe ($\S\ref{sec:56Ni}$), prompt gamma-ray emission ($\S\ref{sec:GRB}$), GRBs with complex light curves such as long temporal ``gaps" ($\S\ref{sec:precursors}$), binary NS mergers ($\S\ref{sec:BNS}$), and SLSNe ($\S\ref{sec:SLSNe}$).
In $\S\ref{sec:conclusions}$, we briefly summarize our conclusions.

\section{Magnetar Model with Accretion}
\label{sec:model}
\subsection{Fall-Back Accretion}

The quantity and return rate of mass to the central neutron star following the core collapse of a massive star depends sensitively on the energy, asymmetry, and detailed mechanism of the explosion.  Although large quantities of fall-back are rare in one-dimensional supernova models which undergo successful explosions (e.g.~\citealt{Ugliano+12,Sukhbold+16}), fall-back in the equatorial plane defined by the angular momentum of the star may be more likely for polar-asymmetric MHD-powered explosions likely responsible for the birth of millisecond magnetars (e.g.~\citealt{Mosta+14}).  Mass accretion following the merger of two NSs also depends sensitively on the properties of the initial binary, such as the mass ratio of its components (e.g.~\citealt{Hotokezaka+13}).  For simplicity, we assume a generic time-dependence for the mass fall-back rate of the form
\be
\dot{M}(t) = \frac{2}{3}\frac{M_{\rm fb}}{t_{\rm fb}}\frac{1}{(1 +t/t_{\rm fb})^{5/3}},
\label{eq:mdotfb}
\ee
where $M_{\rm fb}$ is the total quantity of returning mass and $t_{\rm fb} \sim (G\bar{\rho})^{-1/2}$ is the characteristic fall-back time which depends on the mean density $\bar{\rho}$ of the layer of progenitor star contributing to $M_{\rm fb}$.  For example, the region just outside of the iron core ($\bar{\rho} \lesssim 10^{8}$ g cm$^{-3}$) will fall back first, on a timescale $t_{\rm fb} \sim $ seconds.  The envelope of a compact stripped-envelope star ($\bar{\rho} \sim 1-100$ g cm$^{-3}$) will return on a time $t_{\rm fb} \sim 100-1000$ s.  More extended stars like blue supergiants ($\bar{\rho} \sim 10^{-5}-10^{-3}$ g cm$^{-3}$) will have $t_{\rm fb} \sim 10^{5}-10^{6}$ s, while red supergiants ($\bar{\rho} \sim 10^{-7}$ g cm$^{-3}$) will have $t_{\rm fb} \sim 10^{7}$ s.  A NS of initial mass $M_{\rm ns} \approx 1.4M_{\odot}$ can accrete at most $M_{\rm max} - M_{\rm ns} \approx 0.6-0.8M_{\odot}$ before collapsing into a black hole (depending on the uncertain maximum mass of a NS, $M_{\rm max} \gtrsim 2M_{\odot}$).  Thus, depending on the progenitor star, we should consider accretion rates in the range $\dot{M} \sim M_{\rm fb}/t_{\rm fb} \sim 10^{-8}-0.1M_{\odot}$ s$^{-1}$ acting over characteristic timescales ranging from seconds to months.  Although the mass fall-back rate decays as $\propto t^{-5/3}$ at late times, the accretion rate reaching small radii may decay shallower in time $\dot{M} \propto t^{-4/3}$ if instead the dominant source of mass feeding is a viscously-spreading accretion disk (\citealt{Metzger+08c}). 

Fall-back material with specific angular momentum $j_{\rm fb} = (GM_{\rm ns}R_{\rm fb})^{1/2}$ will circularize into an centrifugally-supported torus outside the NS at the radius $\sim R_{\rm fb}$ (we neglect possible complications arising from thermal pressure support behind the accretion shock; \citealt{Milosavljevic+12}).  For typical radial angular momentum profiles of the progenitor star, the shell-averaged value of $j_{\rm fb}(r)$ increases with radius (\citealt{Heger+05}) and thus, as different layers of the star fall in, $R_{\rm fb}$ should increase monotonically with time (e.g.~\citealt{Kumar+08}).  For simplicity in what follows we assume that $R_{\rm fb}$ is roughly constant in time, motivated by the fact that most of the mass-fall back (arriving on the characteristic timescale $\sim t_{\rm fb}$) originates from a given stellar layer with a roughly constant value of $j_{\rm fb}$.  We also assume that $R_{\rm fb}$ exceeds other critical radii in the problem, such as the light cylinder, co-rotation, and Alfv\'en radii, all of which are typically $\lesssim 100$ km for magnetar properties and accretion rates of interest (Table \ref{table:radii}; $\S\ref{sec:radii}$).  These assumptions can be generalized in future work, for instance to simultaneously account for the evolution of a viscously-spreading disk.

\begin{table*}
\begin{center}
\caption{Critical Radii}
\begin{tabular}{@{}ccc@{}}

\hline
\hline

Radius & Symbol & Value (km)\\

\hline
Neutron Star & $R_{\rm ns}$ & 12 \\
Light Cylinder & $R_{\rm lc}$ & 47.7$P_{\rm ms}$ \\
Co-Rotation & $R_{\rm c}$ & 16.8$P_{\rm ms}^{2/3}M_{1.4}^{1/3}$ \\
Alfv\'en & $R_{\rm m}$ & 22.2$B_{15}^{4/7}\dot{M}_{-2}^{-2/7}M_{1.4}^{-1/7}$ \\
\hline
\hline
\label{table:radii}

\end{tabular}
\end{center}
\end{table*}

\subsection{Critical Radii and Accretion Rates}
\label{sec:radii}

The accretion disk extends inwards from its circularization point $R_{\rm fb}$, feeding matter towards the NS at the rate $\dot{M}$.  We neglect mass loss due to winds from the disk, as this can be accounted for by rescaling the effective value of the accreted mass $M_{\rm fb}$ in equation (\ref{eq:mdotfb}).  The disk is truncated near the Alfv\'en radius $R_{\rm m}$, as interior to this point the NS magnetosphere has a dominant influence over the accretion flow, directing the inflowing gas along field lines close to the edge of the polar funnel. 

The magnetar is assumed to possess an equatorial surface dipole field strength of $B_{\rm d} = B_{15}\times 10^{15}$ G and magnetic moment $\mu = B_{\rm d}R_{\rm ns}^{3}$ which is aligned with the rotation axis, where $R_{\rm ns} = 12$ km is the NS radius.  For an accretion rate $\dot{M} = \dot{M}_{-2}\times 10^{-2}M_{\odot}$s$^{-1}$, the Alfv\'en radius can be estimated as (e.g.~\citealt{Ghosh&Lamb78})
\begin{eqnarray}
R_{\rm m} &=& \left(\frac{3B_{\rm d}^{2} \, R_{\rm ns}^{6}}{2\dot{M}\sqrt{GM_{\rm ns}}}\right)^{2/7} \nonumber \\
&\approx& 22.2\,B_{15}^{4/7}\dot{M}_{-2}^{-2/7}M_{1.4}^{-1/7}\,{\rm km},
\label{eq:RA}
\end{eqnarray}
where $M_{\rm ns} = 1.4 M_{1.4}M_{\odot}$ is the NS mass.  

The central magnetar rotates with an angular frequency $\Omega = 2\pi/P$ and spin period $P = P_{\rm ms}\times 1$ ms.  Two additional key radii are the light cylinder radius,
\be  R_{\rm lc}  = c/\Omega = 47.7P_{\rm ms}\,{\rm km} \ee
and the co-rotation radius
\be
R_{\rm c} = \left(\frac{GM_{\rm ns}}{\Omega^{2}}\right)^{1/3} \simeq 16.8 M_{1.4}^{1/3}P_{\rm ms}^{2/3}\,{\rm km}.
\ee
These correspond, respectively, to the locations at which the azimuthal velocity of matter co-rotating with the magnetar equals the speed of light and the Keplerian velocity.  These critical radii are summarized in Table \ref{table:radii}.

\begin{figure*}[!t]
%\includegraphics[width=0.7\textwidth]{e}
%\hspace{0.0cm}
\includegraphics[width=1.0\textwidth]{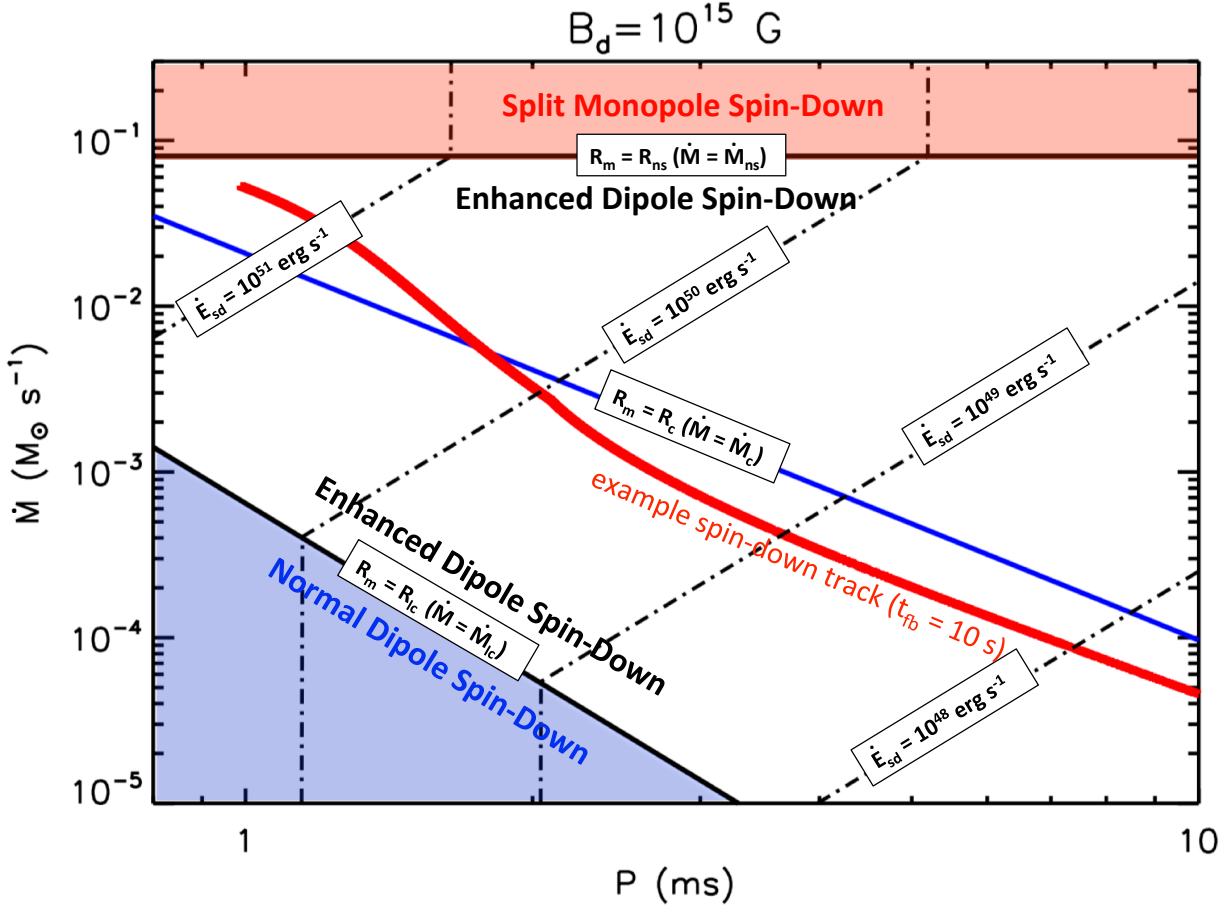}
\vspace{-0.4cm}
\caption{\footnotesize Regimes of magnetar spin-down evolution in the parameter space of mass accretion rate $\dot{M}$ and spin period $P$, calculated for a magnetar of radius $R_{\rm ns} = 12$ km, mass $M_{\rm ns} = 1.4M_{\odot}$, and surface dipole magnetic field strength $B_{\rm d} = 10^{15}$ G.
  Solid lines show accretion rates $\dot{M} = \dot{M}_{\rm ns}$ (eq.~\ref{eq:Mdotns}), $\dot{M}_{\rm c}$ (blue; eq.~\ref{eq:Mdotc}), $\dot{M}_{\rm lc}$ (black, eq.~\ref{eq:Mdotlc}).  The condition $\dot{M} = \dot{M}_{\rm c}$ approximately corresponds to the equilibrium spin period achieved if accretion spin-up is balanced by torques from the star-magnetosphere interaction (eq.~\ref{eq:Peq}).  Dot-dashed lines show contours of constant spin-down luminosity $\dot{E}_{\rm sd}$ (eq.~\ref{eq:edotsdcases}).  A red solid line shows an example evolutionary track for an accreting magnetar ($t_{\rm fb} = 10$ s case shown in Fig.~\ref{fig:results}).  Though we show results for a particular value of $B_{\rm d} = 10^{15}$ G, all of the critical accretion rates scale as $\dot{M} \propto B_{\rm d}^{2}$.}
\label{fig:MdotP}
\end{figure*}

As we describe below, the rate of magnetar spin-down due to angular momentum loss from a magnetized wind is enhanced when the Alfv\'en radius resides interior to the light cylinder.  This condition ($R_{\rm m} < R_{\rm lc}$) is satisfied above a critical accretion rate of
\be
\dot{M}_{\rm lc} \simeq 6.9\times 10^{-4} B_{15}^{2}P_{\rm ms}^{-7/2}M_{1.4}^{-1/2}M_{\odot}\,{\rm s^{-1}}, R_{\rm m} = R_{\rm lc}, 
\label{eq:Mdotlc}
\ee
or, correspondingly, below a spin period of
\be
P_{\rm lc} \simeq 0.47 B_{15}^{4/7}\dot{M}_{-2}^{-2/7}M_{1.4}^{-1/7} {\rm ms} .
\ee
The enhancement of the spin-down rate saturates once $R_{\rm m}$ is pushed all the way down to the NS surface, a condition that requires an even higher accretion rate of
\be
\dot{M}_{\rm ns} \simeq 0.086 B_{15}^{2}M_{1.4}^{-1/2}M_{\odot}\,{\rm s^{-1}}, R_{\rm m} = R_{\rm ns},
\label{eq:Mdotns} 
\ee
Whether matter accretes freely onto the magnetar, or whether the system may enter the more complicated ``propeller regime" (e.g.~\citealt{Romanova+03}), depends on the location of the co-rotation radius relative to the Alfv\'en radius.  $R_{\rm m}$ exceeds $R_{\rm c}$ for accretion rates above a critical value of
\be
\dot{M}_{\rm c} \simeq 0.026 B_{15}^{2}P_{\rm ms}^{-7/3}M_{1.4}^{-5/3}M_{\odot}\,{\rm s^{-1}}, R_{\rm m} = R_{\rm c} 
\label{eq:Mdotc}
\ee
or, correspondingly, below a spin period of
\be
P_{\rm c} \simeq 1.52 B_{15}^{6/7}\dot{M}_{-2}^{-3/7}M_{1.4}^{-5/7}{\rm ms}.
\ee
These critical values of the accretion rate are summarized in Fig.~\ref{fig:MdotP} as a function of spin period for $B_{\rm d} = 10^{15}$ G.  The hierarchy $\dot{M}_{\rm lc} \lesssim \dot{M}_{\rm c} \lesssim \dot{M}_{\rm ns}$ is preserved for all values of $B_{\rm d}$ and thus is generic also to the lower values of $\dot{M}$ and $B_{\rm d}$ more relevant to  SLSNe.

\subsection{Magnetar Spin Evolution}

\begin{figure}[!t]
\includegraphics[width=0.5\textwidth]{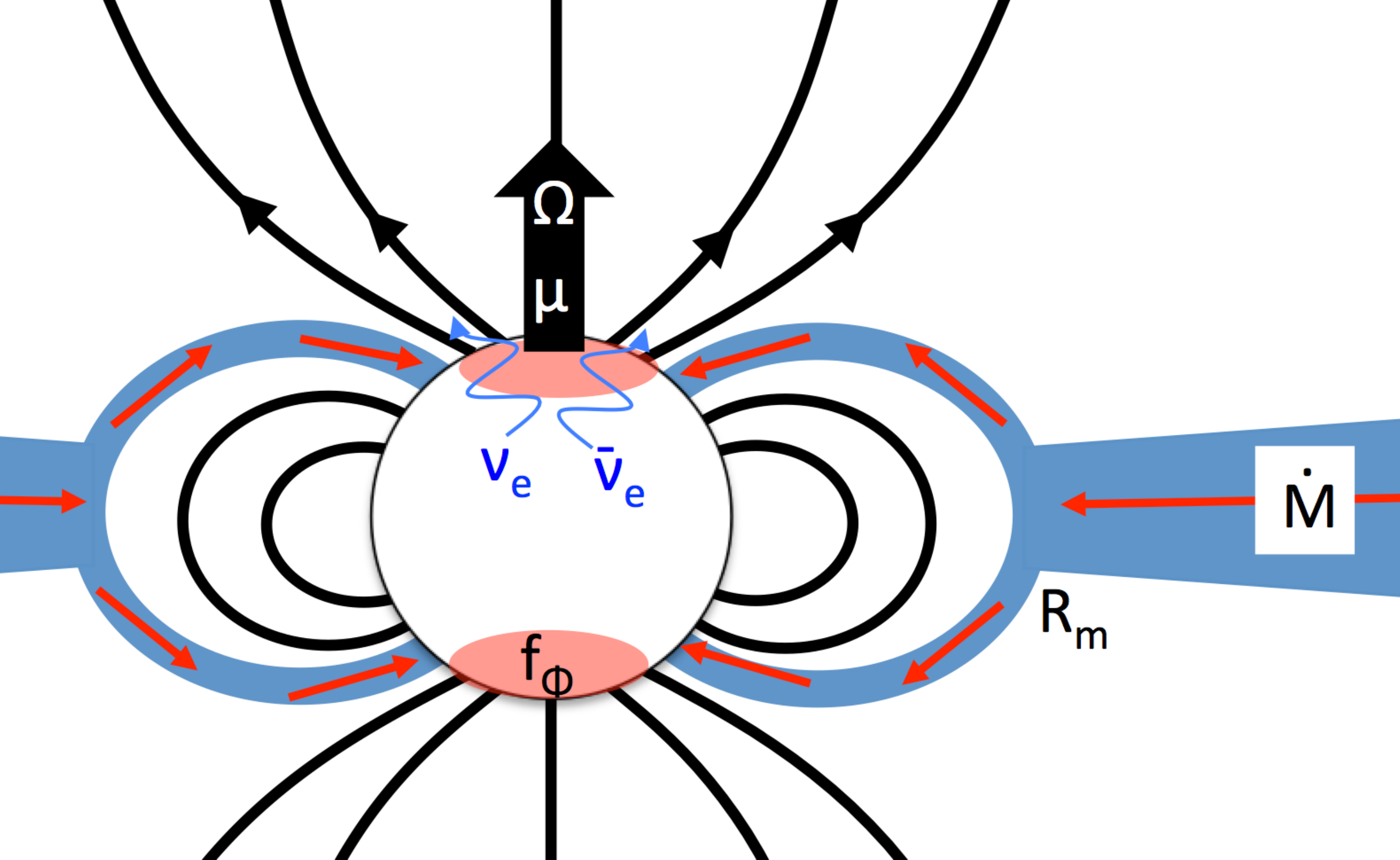}
\hspace{0.0cm}
\vspace{-0.4cm}
\caption{\footnotesize Schematic diagram of an accreting proto-magnetar and its magnetized relativistic outflow.  The polar cap of open magnetic flux (shown in red) controls the spin-down rate of the magnetar and pollutes the jet with baryons.  The rate of baryonic mass loss from the NS surface is controlled by neutrino heating in the proto-magnetar atmosphere ($\nu_{e}+n \rightarrow p + e^{-}$; $\bar{\nu}_{e} + p \rightarrow n + e^{+}$).  The neutrinos arise both from the NS interior (until the NS becomes optically thin at times $t \lesssim t_{\rm thin}$) and are produced in the magnetic accretion column, as shown in blue.}
\label{fig:cartoon}
\end{figure}

The rotational energy of the magnetar is given by
\be
E_{\rm rot} = \frac{1}{2}I\Omega^{2} \simeq 2.6\times 10^{52}\,M_{1.4}^{3/2}P_{\rm ms}^{-2}\,{\rm ergs},
\label{eq:Erot}
\ee
where $I$ is the NS moment of inertia, which we approximate as $I \simeq 1.3\times 10^{45}M_{1.4}^{3/2}$ g cm$^{2}$ (Fig.~1 of \citealt{Lattimer&Schutz05}).  Maximum rotation energies of $E_{\rm rot} \approx 3\times 10^{52}$ ergs are typical for NSs of mass 1.4$M_{\odot}$ rotating near the break-up period of $P \simeq 1$ ms, but reservoirs up to $E_{\rm rot} \approx 10^{53}$ ergs are possible for more massive NSs with $M_{\rm ns} \gtrsim 2M_{\odot}$ (\citealt{Metzger+15}).

Following the formation of a magnetar, its angular momentum $J = I \Omega$ evolves in time according to
\be
\frac{dJ}{dt} = \dot{J}_{\rm acc} - \dot{J}_{\rm sd},
\label{eq:jdot}
\ee
where $\dot{J}_{\rm sd} = \dot{E}_{\rm sd}/\Omega$ accounts for losses due to the Poynting flux of the magnetized wind along the polar directions, and $\dot{J}_{\rm acc}$ accounts for gains or losses due to the torque from the accretion disk (see Fig.~\ref{fig:cartoon}).  We now discuss each of these processes in detail.

The magnetar loses rotational energy to a polar Poynting flux-dominated wind at a rate
\be
\dot{E}_{\rm sd} = \left(\frac{f_{\Phi}}{f_{\Phi, \rm lc}}\right)^{2}\frac{\mu^{2}\Omega^{4}}{c^{3}},
\label{eq:edotsd}
\ee
where we have assumed that the magnetic dipole and rotational axes are aligned (see \citealt{Margalit+18} for generalization to the misaligned case).  Here $f_{\Phi} = \int_{0}^{\theta_{\rm lc}}\sin \theta d\theta \approx \theta_{\rm lc}^{2}/2$ is the fraction of the magnetar surface threaded by open magnetic flux and $\theta_{\rm lc}$ is the latitude from the pole of the last closed field line  (e.g.~\citealt{Bucciantini+06}).  For an isolated pulsar wind, the closed field lines are typically assumed to extend to the light cylinder $R_{\rm lc}$, corresponding to $\theta_{\rm lc} \approx \sin^{-1}(R_{\rm ns}/R_{\rm lc})^{1/2} \approx (R_{\rm ns}/R_{\rm lc})^{1/2}$ (e.g.~\citealt{Contopoulos+99,Spitkovsky06}), thus providing a minimum open flux of $f_{\Phi,\rm lc} = R_{\rm ns}/2R_{\rm lc} \approx 0.13P_{\rm ms}^{-1}$.  In this limit ($f_{\Phi} = f_{\Phi,\rm lc}$), one recovers from equation (\ref{eq:edotsd}) the normal magnetic dipole spin-down rate $\dot{E}_{\rm sd} \propto \Omega^{4}$ of the force-free wind.

By contrast, for an accreting NS with $R_{\rm m} \lesssim R_{\rm lc}$, closed field lines are truncated where the disk begins, in which case $\theta_{\rm lc}  \approx (R_{\rm ns}/R_{\rm m})^{1/2}$ and $\dot{E}_{\rm sd}$ is thus enhanced over the standard dipole rate by a factor of $(R_{\rm m}/R_{\rm lc})^{-2} > 1$ (\citealt{Parfrey+16}).  The maximum spin-down rate is achieved for the maximum open flux $R_{\rm m} = R_{\rm ns}$, i.e. the limit of a ``split monopole" field geometry.  Combining results,  the wind energy loss rate is given by
\begin{eqnarray} &&\dot{E}_{\rm sd} = \left\{
\begin{array}{lr} 
\frac{\mu^{2}\Omega^{4}}{c^{3}}\frac{R_{\rm lc}^{2}}{R_{\rm ns}^{2}}, & \dot{M} \gtrsim \dot{M}_{\rm ns}, \\
\frac{\mu^{2}\Omega^{4}}{c^{3}}\frac{R_{\rm lc}^{2}}{R_{\rm m}^{2}} , & \dot{M}_{\rm lc} \lesssim \dot{M} \lesssim \dot{M}_{\rm ns}, \\
\frac{\mu^{2}\Omega^{4}}{c^{3}}, &
\dot{M} \lesssim \dot{M}_{\rm lc}\\
\end{array}
\label{eq:edotsdcases}
\right. , \nonumber \\
 &\approx& \left\{
\begin{array}{lr}  2.7\times 10^{51}B_{15}^{2}P_{\rm ms}^{-2}\,{\rm ergs\,s^{-1}}, & \dot{M} \gtrsim \dot{M}_{\rm ns}, \\
\ 8\times 10^{50}B_{15}^{6/7}P_{\rm ms}^{-2}\dot{M}_{-2}^{4/7}M_{1.4}^{2/7}\,{\rm ergs\,s^{-1}}, & \dot{M}_{\rm lc} \lesssim \dot{M} \lesssim \dot{M}_{\rm ns}, \\
1.7\times 10^{50}B_{15}^{2}P_{\rm ms}^{-4}\,{\rm ergs\,s^{-1}}, &
\dot{M} \lesssim \dot{M}_{\rm lc}\\
\end{array}
\label{eq:edotsdcases}
\right. \nonumber \\,
\end{eqnarray}
Contours of fixed spin-down power are shown as dashed lines in Fig.~\ref{fig:MdotP}.

A characteristic wind spin-down time is defined according to
\begin{eqnarray}
&&t_{\rm sd} \equiv \frac{E_{\rm rot}}{\dot{E}_{\rm sd}} \nonumber \\
&=& \left\{
\begin{array}{lr} 
t_{\rm sd,1} \simeq 9 B_{15}^{-2}M_{1.4}^{3/2}\,{\rm s} & \dot{M} \gtrsim \dot{M}_{\rm ns},  \\
t_{\rm sd,2} \simeq 33 B_{15}^{-6/7}\dot{M}_{-2}^{-4/7}M_{1.4}^{17/14}\,{\rm s}, & \dot{M}_{\rm lc} \lesssim \dot{M} \lesssim \dot{M}_{\rm ns}, \\ t_{\rm sd,3} \simeq 150B_{15}^{-2}P_{\rm ms}^{2}M_{1.4}^{3/2}\,{\rm s}, &\dot{M} \lesssim \dot{M}_{\rm lc} \\
\end{array}
\label{eq:tsd}
\right. \nonumber \\
\end{eqnarray}
When the accretion of mass or angular momentum from the disk can be neglected ($\dot{J}_{\rm acc} = 0$, $dM/dt = 0$), and the accretion rate is constant in time, the magnetar spin-down $dE_{\rm rot}/dt = -\dot{E}_{\rm sd}$ exhibits a simple analytic solution in each of these regimes,
\be
\dot{E}_{\rm sd} = \left\{
\begin{array}{lr} 
\frac{E_{\rm rot,0}}{t_{\rm sd,1}}e^{-t/t_{\rm sd,1}} & \dot{M} \gtrsim \dot{M}_{\rm ns},\\
\frac{E_{\rm rot,0}}{t_{\rm sd,2}}e^{-t/t_{\rm sd,2}} & \dot{M}_{\rm lc} \lesssim \dot{M} \lesssim \dot{M}_{\rm ns}, \\
\frac{E_{\rm rot,0}}{t_{\rm sd,3}}(1 + t/t_{\rm sd,3})^{-2} & \dot{M} \lesssim \dot{M}_{\rm lc},\\
\end{array}
\label{eq:edotsdevo}
\right. ,
\ee
where $E_{\rm rot,0}$ is the initial rotational energy.  Because both $t_{\rm sd,1}$ and $t_{\rm sd,2}$ are independent of the spin period, in these regimes the spin-down power decays as an exponential instead of the usual late-time power-law decay $\propto t^{-2}$ for dipole spin-down.

In addition to the losses from the magnetized wind, the magnetar can exchange angular momentum with the accretion disk at the rate (e.g.~\citealt{Piro&Ott11})
\be
\dot{J}_{\rm acc} = \left\{
\begin{array}{lr} \dot{M}(GM_{\rm ns}R_{\rm ns})^{1/2}(1-\Omega/\Omega_{\rm K}), & \dot{M} \gtrsim \dot{M}_{\rm ns}  \\
\dot{M}(GM_{\rm ns}R_{\rm m})^{1/2}n(\omega)  & \dot{M} \lesssim \dot{M}_{\rm ns} \\
\end{array}
\label{eq:jdotacc}
\right. .
\ee
Here the $(1-\Omega/\Omega_{\rm K})$ factor in eq.~\ref{eq:jdotacc} prevents the NS from gaining additional angular momentum once it is rotating near the centrifugal break-up velocity\footnote{Physically, this limit could be enforced by centrifugally-driven mass-loss or efficient gravitational wave losses induced non-axisymmetric instabilities which set in at high $T/|W|$ approaching the break-up threshold (e.g.~\citealt{Lai&Shapiro95}).} at $\Omega \approx \Omega_{\rm K} = (GM_{\rm ns}/R_{\rm ns}^{3})^{1/2}$.

Whether the disk spins up ($\dot{J}_{\rm acc} > 0$) or spins down ($\dot{J}_{\rm acc} < 0$) the magnetar will in general depend on the fastness parameter $\omega \equiv (R_{\rm m}/R_{\rm c})^{3/2}$.  The precise way that the torque changes around $\omega \approx 1$ is poorly understood and remains a matter of debate in the literature (e.g.~\citealt{DAngelo&Spruit12}).  We explore two models for $n(\omega)$.  First, as our fiducial case, we consider the prescription 
\be 
n(\omega) = 1-\omega , \,\,\,\,\,\,\,\text{Piro \& Ott}
\label{eq:PiroOtt}
\ee
of \citet{Piro&Ott11}, which allows for the loss of angular momentum from the NS in the so-called ``propeller regime" $\omega \gtrsim 1$.  On the other hand, \citet{Parfrey+16} argue for a minimal coupling between the NS magnetic field and the disk matter in the nominal propeller regime and instead take 
\be
n(\omega) = \left\{
\begin{array}{lr} 1, &
\omega<1 \\
0 & \omega \ge 1. \\
\end{array}
\,\,\,\,\,\,\,\,\text{Parfrey}
\label{eq:Parfrey}
\right. 
\ee
As we show below, these different prescriptions can lead to qualitative differences in the magnetar evolution, such as whether an equilibrium spin period is always achieved.

Accretion also causes the magnetar to grow in mass.  How efficiently the star actually accepts the matter being fed from the disk depends on uncertain factors, such as whether the polar accretion column is able to cool through neutrinos and settle on the NS surface \citep{Piro&Ott11}.  Another uncertainty is the efficiency with which matter accretes in the propeller regime \citep{Romanova+04}; some X-ray binaries believed to be accreting in the propeller regime nevertheless show X-ray emission from accretion (e.g.~\citealt{Gungor+17}).  We assume that growth of the NS mass occurs at the rate
\be
\frac{dM_{\rm ns}}{dt}  = \left\{
\begin{array}{lr} \dot{M}, &
\dot{M} \gtrsim \dot{M}_{\rm c} \\
f_{\rm acc}\dot{M} & \dot{M} \lesssim \dot{M}_{\rm c} \\
\end{array}
\label{eq:mdotacc}
\right. ,
\ee
where $f_{\rm acc} < 1$ is the accretion efficiency in the propeller regime.  Though we adopt $f_{\rm acc} = 0$ in what follows, this is easily generalizable and our main conclusions are not sensitive to this assumption.

An accreting magnetar can reach a approximate equilibrium for which $dJ/dt \approx 0$, either exactly or in a time-averaged sense, depending on the adopted prescription for angular momentum loss in the propeller regime, $n(\omega)$.  Accretion spin-up ($\dot{J}_{\rm acc} > 0$) can be balanced either by wind spin-down ($\dot{J}_{\rm sd} = \dot{E}_{\rm sd}/\Omega < 1$) or propeller spin-down ($\dot{J}_{\rm acc} < 0$; in the \citealt{Piro&Ott11} prescription).  This equilibrium occurs for $\omega \approx 1$, corresponding to a spin period of
\be
P_{\rm eq} \underset{\omega = 1}= P_{\rm c} \simeq 1.52 B_{15}^{6/7}\dot{M}_{-2}^{-3/7}M_{1.4}^{-5/7}{\rm ms}.
\label{eq:Peq}
\ee
The timescale needed to maintain equilibrium can be estimated as
\be
\tau_{\rm eq} \equiv \frac{I \Omega_{\rm eq}}{\dot{M}(GM_{\rm ns}R_{\rm m})^{1/2}} \approx  13.2\,B_{15}^{-8/7}\dot{M}_{-2}^{-3/7}M_{1.4}^{16/7}\,{\rm s},
\label{eq:taueq}
\ee
where $\Omega_{\rm eq} = 2\pi/P_{\rm eq}$.

 This equilibrium is in some sense more ``robust" for the \citet{Piro&Ott11} coupling prescription, because if $\omega$ becomes $\gtrsim 1$, then the disk torque becomes negative (propeller spin-down), driving $\omega$ back to $\simeq 1$.  As long as the mass accretion rate evolves relatively slowly compared to $\tau_{\rm eq}$ then $P \approx P_{\rm eq}$ will be maintained.  

Though a similar equilibrium condition (\ref{eq:Peq}) may be achieved also for the \citet{Parfrey+16}  prescription (eq.~\ref{eq:Parfrey}), in this case if $\omega$ evolves to become $\gtrsim 1$ then the disk no longer exerts a torque on the star.  As long as the accretion rate evolves slowly, then spin-down of the star will drive $\omega \lesssim 1$ again, temporarily restoring $\omega \approx 1$ and resulting in a time-average equilibrium with $P \approx P_{\rm eq}$.  However, if instead the accretion rate decreases  rapidly compared to the spin-down time of the isolated magnetar (i.e.~$t_{\rm fb} \ll t_{\rm sd,3}$), then the star can decouple permanently from the disk and spin-down will proceed independently of an subsequent mass fall-back.

\section{Results}
\label{sec:results}

\begin{figure*}[!t]
\centering
\includegraphics[width=0.9\textwidth]{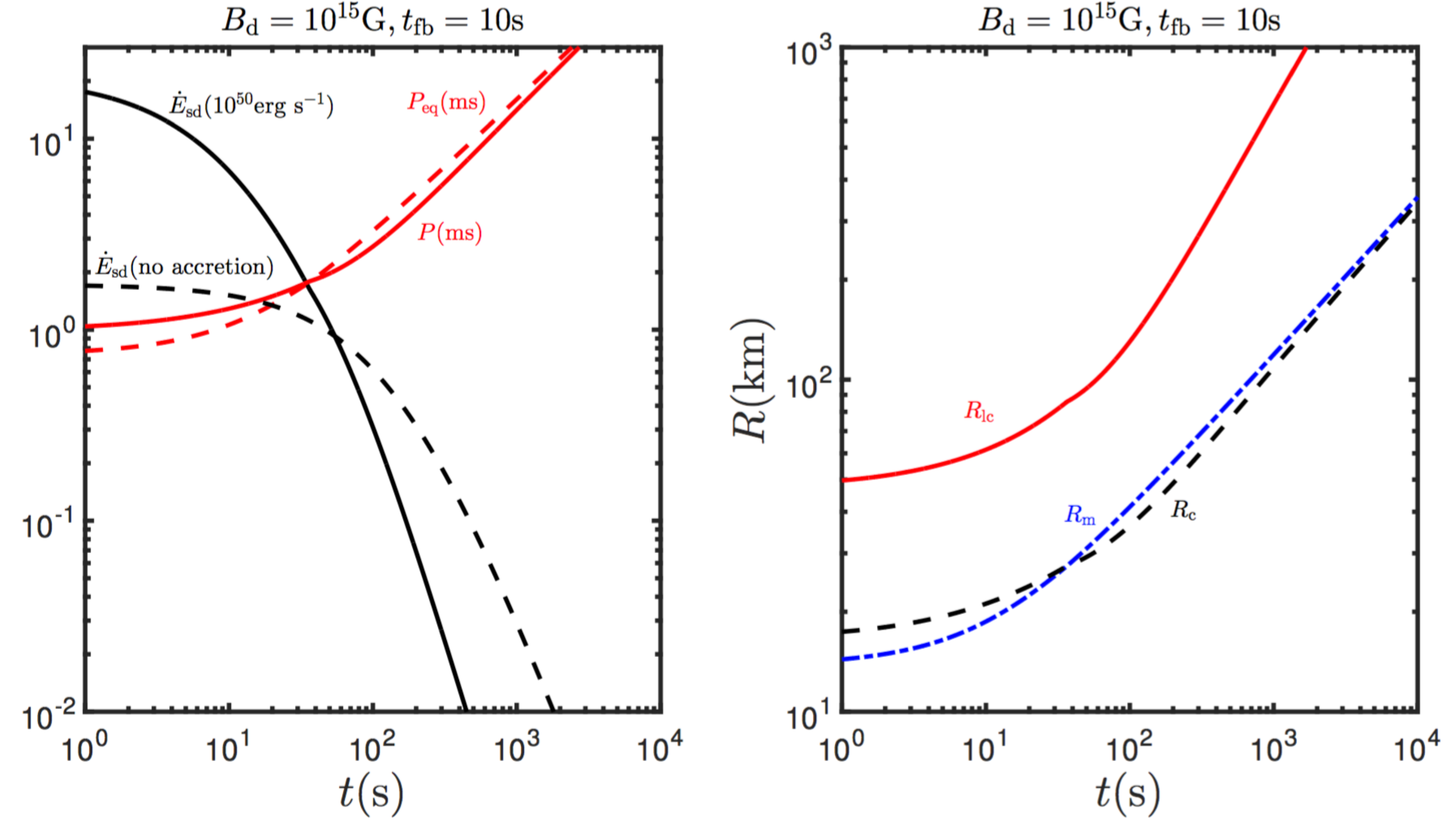}\\
\includegraphics[width=0.9\textwidth]{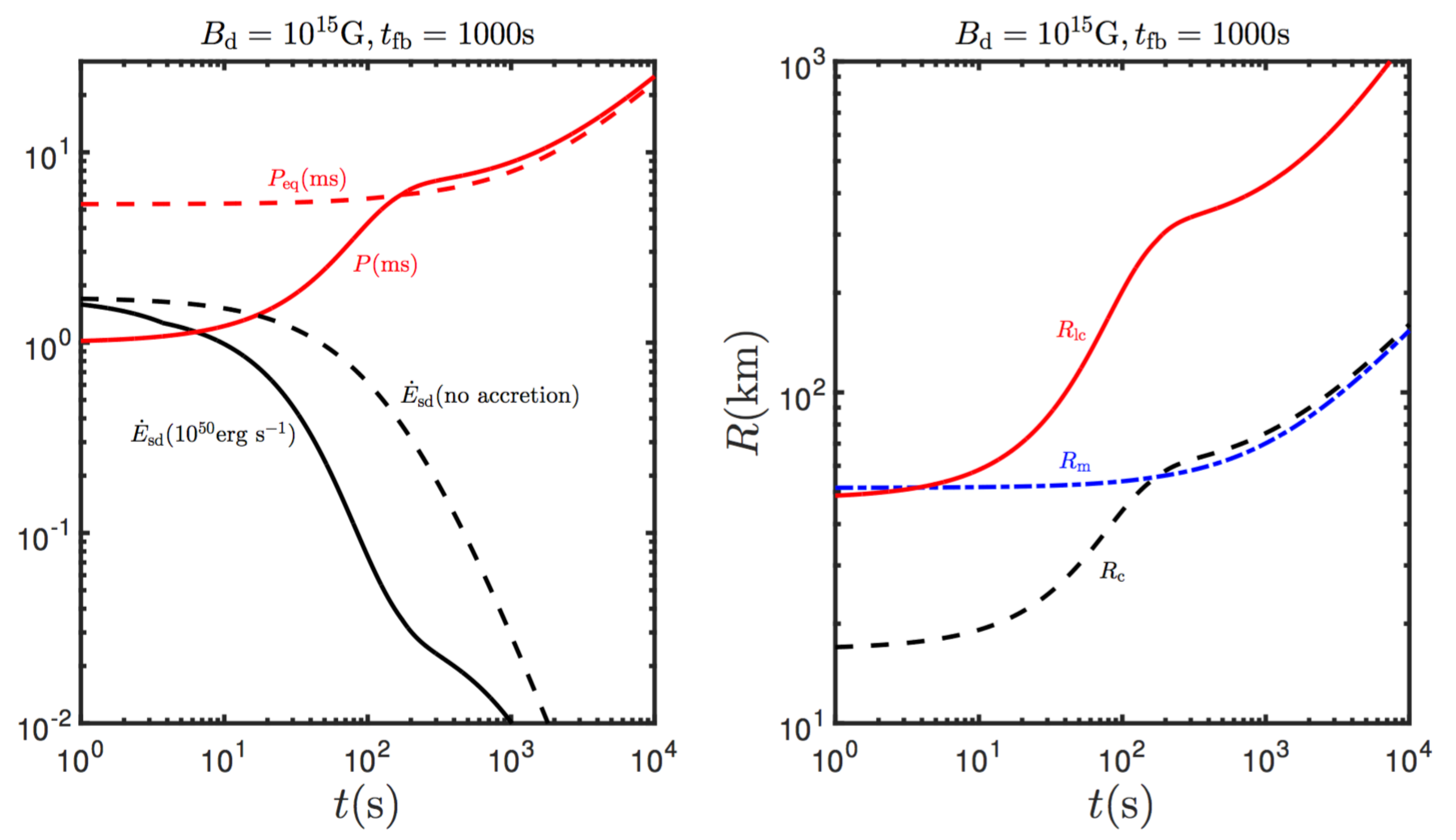}
\caption{\footnotesize Spin down evolution of an accreting magnetar with surface dipole field strength of $B_{\rm d} = 10^{15}$ G and initial spin period $P_{0} = 1$ ms, assuming mass fall-back with $M_{\rm fb} = 0.8M_{\odot}$ over a timescale $t_{\rm fb} = 10$ s (top panels) and $t_{\rm fb} = 10^{3}$ s (bottom panels).  The left panels show the spin-down luminosity $\dot{E}_{\rm sd}$ (black solid curve), compared to its equivalent evolution without fall-back accretion (black dashed curve).  Also shown is the magnetar spin period $P$ (solid red curve) compared to its value in equilibrium $P_{\rm eq} = P_{\rm c}$ (dashed red curve) at which $R_{\rm m} \approx R_{\rm c}$ (eq.~\ref{eq:Peq}).  The right panels shows several critical radii: light cylinder $R_{\rm lc}$ (solid red curve), Alfv\'en $R_{\rm m}$ (dotted blue curve), and co-rotation $R_{\rm c}$ (dashed black curve).  These models are calculated using the \citet{Piro&Ott11} prescription for the $\dot{J}$ coupling between the accretion disk and NS magnetosphere (eq.~\ref{eq:PiroOtt}).  }
\label{fig:results}
\end{figure*}

\begin{figure*}[!t]
\centering
\includegraphics[width=0.9\textwidth]{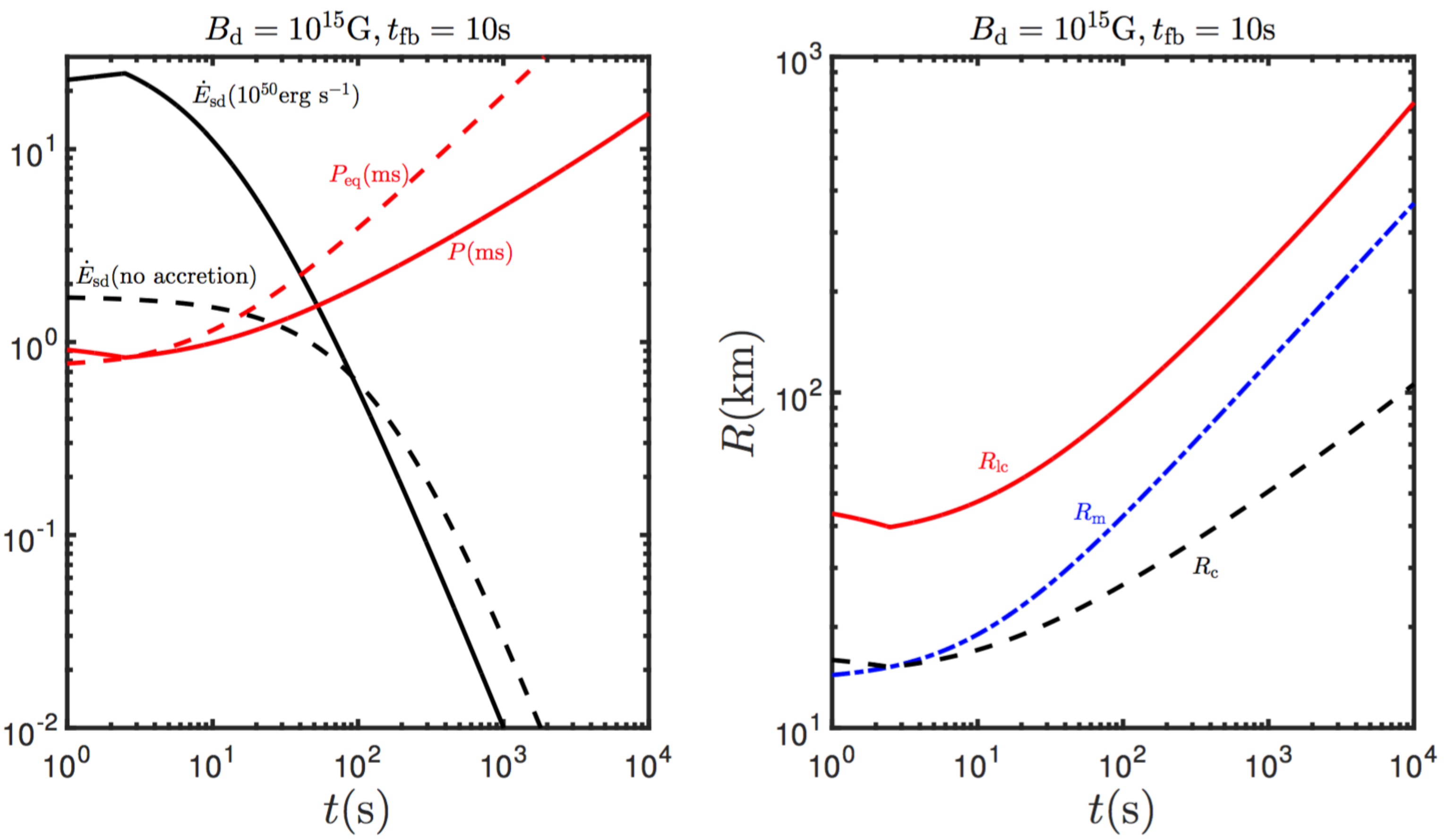}\\
\includegraphics[width=0.9\textwidth]{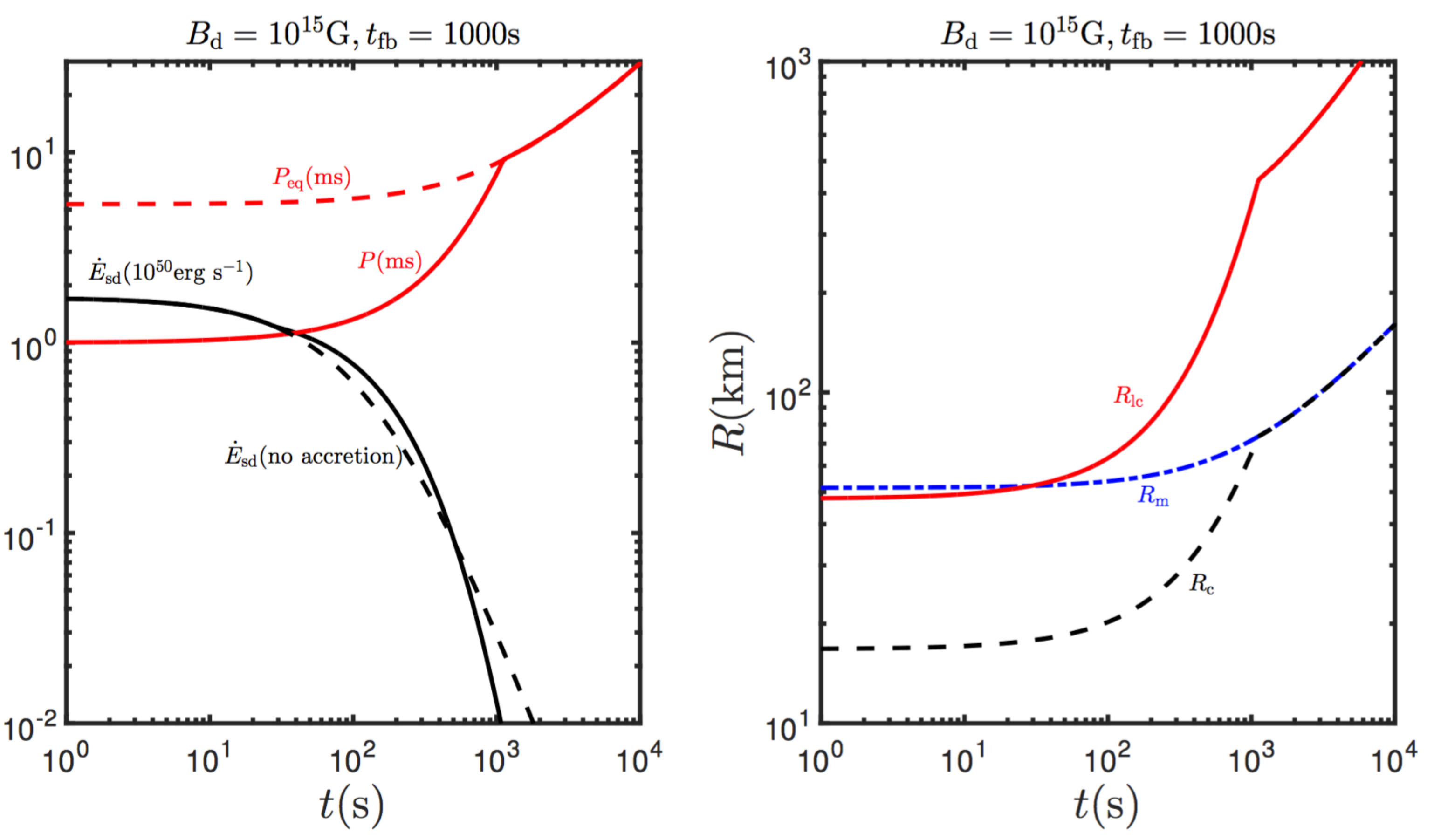}
\caption{\footnotesize Same as Fig. \ref{fig:results}, but assuming zero $\dot{J}$ coupling between the accretion disk and the NS magnetosphere in the propeller regime (\citealt{Parfrey+16}; eq.~\ref{eq:Parfrey}). }
\label{fig:results2}
\end{figure*}
\subsection{Spin-Down Evolution}
\label{sec:spindown}

We solve equations (\ref{eq:jdot}), (\ref{eq:mdotacc}) for the evolution of the magnetar spin period $P(t)$ and the spin-down power $\dot{E}_{\rm sd}(t)$ given the initial properties of the magnetar ($B_{\rm d}$, birth spin period $P_{0}$) and the externally imposed mass fall-back evolution $\dot{M}(t)$ (eq.~\ref{eq:mdotfb}). Because we are interested in exploring the maximal effects of accretion, we fix the total fall-back mass at the maximum value $M_{\rm fb} = 0.8 M_{\odot}$ needed to avoid gravitational collapse to a black hole and explore the dependence of the result on the fall-back time, $t_{\rm fb}$.  

Figure \ref{fig:results} shows two examples of $\dot{E}_{\rm sd}(t)$ and $P(t)$ for $B_{\rm d} = 10^{15}$ G, $P_{0} = 1$ ms, $f_{\rm acc} = 0$, for different values of the mass fall-back time $t_{\rm fb} = 10$ s (top panel) and $t_{\rm fb} = 10^{3}$ s (bottom panel), employing the \citet{Piro&Ott11} disk-magnetosphere coupling prescription (eq.~\ref{eq:PiroOtt}).  In both cases the spin period achieves the equilibrium value $P_{\rm eq}$ (eq.~\ref{eq:Peq}), but the effects on $\dot{E}_{\rm sd}$ are markedly different. In the high $\dot{M}$ case ($t_{\rm fb} = 10$ s), the initial value of $\dot{E}_{\rm sd}$ exceeds by an order of magnitude the otherwise equivalent case of a non-accreting magnetar (shown for comparison with a black dashed curve) as a result of the enhanced spin-down caused by $R_{\rm m} < R_{\rm lc}$.  

In the low $\dot{M}$ case ($t_{\rm fb} = 10^{3}$ s), the spin equilibrium $P = P_{\rm eq}$ is also rapidly achieved on a timescale $\sim 0.1 t_{\rm fb}$.  However, the evolution starts with $R_{\rm m} \approx R_{\rm lc} \gg R_{\rm c}$, such that, although the initial spin-down luminosity is similar to the isolated magnetar case, angular momentum losses in the propeller regime reduce the total rotational energy which is extracted by the magnetar wind to $\int \dot{E}_{\rm sd}dt \approx 10^{52}$ erg, less than half the value for an otherwise equivalent isolated magnetar with the same birth period.  
  
If the equilibrium $P = P_{\rm eq}$ (eq.~\ref{eq:Peq}) is achieved, then the luminosity of the magnetar wind evolves as
\begin{eqnarray}
\dot{E}_{\rm sd,eq} &\simeq&  \left.\frac{\mu^{2}\Omega^{4}}{c^{3}}\frac{R_{\rm lc}^{2}}{R_{\rm c}^{2}}\right|_{P = P_{\rm eq}} \nonumber \\
&\simeq& 3.44\times 10^{50}B_{15}^{-6/7}\dot{M}_{-2}^{10/7}M_{1.4}^{12/7}\,{\rm ergs\,s^{-1}}.
\label{eq:EdotPeq}
\end{eqnarray}
Taking $\dot{M} = 0.8M_{\odot}/t_{\rm fb}$, the timescale to reach spin equilibrium (eq.~\ref{eq:taueq}) relative to the fall-back time is given by
\be
\frac{\tau_{\rm eq}}{t_{\rm fb}} \approx 0.54B_{15}^{-8/7}(t_{\rm fb}/10\,{\rm s})^{-4/7}M_{1.4}^{16/7},
\ee
while relative to the spin-down time we have
\be
\frac{\tau_{\rm eq}}{t_{\rm sd,2}} \approx 0.55 B_{15}^{-2/7}(t_{\rm fb}/10\,{\rm s})^{-1/7}M_{1.4}^{1/2}.
\ee
For the parameter ranges of interest we have $\tau_{\rm eq} \ll t_{\rm fb}, t_{\rm sd,2}$, indicating that equilibrium will be achieved over timescales relevant to the system evolution.  Furthermore, if $\dot{M} \propto t^{-5/3}$ then $\tau_{\rm eq}/t \propto t^{-2/7}$ at times $t \gg t_{\rm fb}$, such that an equilibrium which is achieved by $t \sim t_{\rm fb}$ will be maintained at all later times despite the declining accretion rate.

If the late-time accretion rate decreases as $\dot{M} \propto t^{-\zeta}$, then the outflow power will approach a power-law decay
\be \dot{E}_{\rm sd} \propto \dot{M}^{10/3} \propto  t^{-10\zeta/3} = \left\{
\begin{array}{lr} t^{-2.38}, &
\zeta = 5/3, \\
t^{-1.90} & \zeta = 4/3. \\
\end{array}
\label{eq:latedecay}
\right. 
\ee 
This decay can be steeper or shallower than the standard $\dot{E}_{\rm sd} \propto t^{-2}$ prediction for isolated dipole spin-down, depending on whether the mass accretion rate is dominated by the mass fall-back rate  ($\zeta = 5/3$) or a viscously-spreading disk ($\zeta = 4/3$).  Also note that this late-time evolution will persist only until the co-rotation radius $R_{\rm c} \propto P^{2/3}$ grows to exceed the circularization radius or the outer edge of the viscously-expanding accretion disk.

Figure \ref{fig:results2} shows for comparison an otherwise identical calculation which instead employs the minimal \citet{Parfrey+16} disk-magnetosphere coupling (eq.~\ref{eq:Parfrey}).  The main qualitative difference, as compared to the case employing the \citet{Piro&Ott11} prescription, is that the equilibrium condition $R_{\rm m} \approx R_{\rm c} (\omega \approx 1)$ no longer necessarily obtains.  If the system begins in the propeller regime ($R_{\rm m} > R_{\rm c}$; $\omega \gtrsim 1$), then whether $\omega \approx 1$ is achieved depends on the timescale over which the mass fall-back rate declines relative to the spin-down time, $t_{\rm sd,3}$.  For the \citet{Parfrey+16} prescription, $\dot{J}_{\rm acc} = 0$ for $\omega > 1$ and spin-down occurs exclusively due to the magnetar wind.  If $t_{\rm fb} \ll t_{\rm sd,3}$ (as in the example shown in the top panel of Fig.~\ref{fig:results2}), then $R_{\rm m}$ increases $\propto t^{10/21}$ while $R_{\rm c}$ grows more slowly in time, such that $\omega > 1$ is maintained and no equilibrium is achieved.  By contrast, if $t_{\rm fb} \gg t_{\rm sd,3}$ (and/or if $\omega < 1$ initially), then the $\omega \approx 1$ condition is achieved, as in the example shown in the bottom panel of Fig.~\ref{fig:results2}.

\begin{figure*}[!t]
\includegraphics[width=0.5\textwidth]{Erot.pdf}
\includegraphics[width=0.5\textwidth]{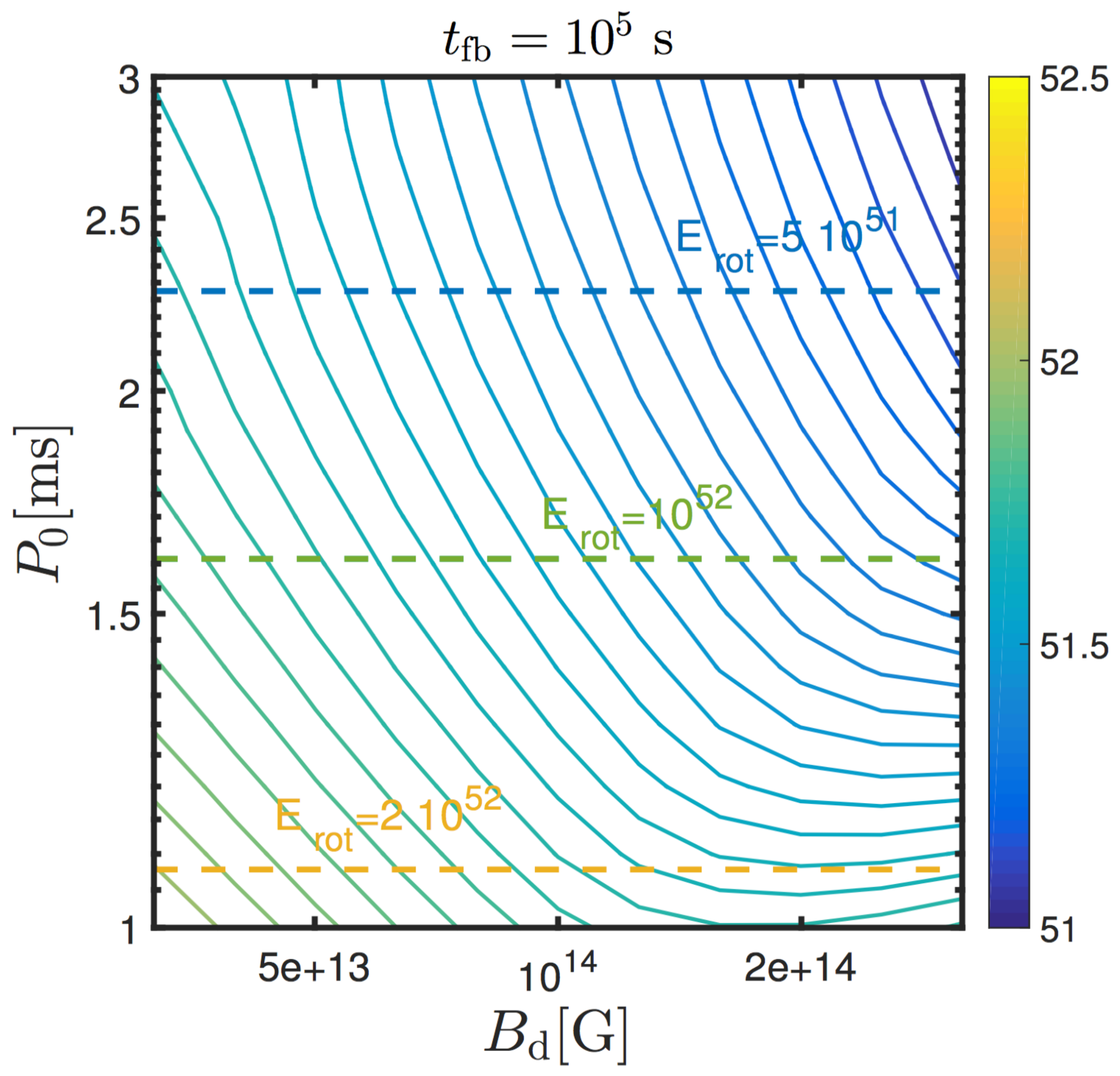}
\caption{\footnotesize Total rotational energy $E = \int \dot{E}_{\rm sd}dt$ extracted in the magnetized wind of a proto-magnetar of initial mass $M_{\rm ns}=1.4M_{\odot}$ as a function of its dipole field strength $B_{\rm d}$ and birth spin period $P_0$, accounting for the effects of mass fall-back of total quantity $M_{\rm fb} = 0.8M_{\odot}$ over a timescale $t_{\rm fb} = 10$ s (left panel) and $t_{\rm fb} = 10^{5}$ s (right panel).  The range of ($B_{\rm d}, P_{0}, t_{\rm fb}$) covered in the two cases correspond to parameters appropriate for magnetars capable of powering long GRB jets and SLSNe, respectively.  The total rotational energy of non-accreting magnetars with the same initial spin period are shown for comparison with dashed lines. }
\label{fig:totalenergy}
\end{figure*}

Figure \ref{fig:totalenergy} shows contours of the total extracted energy $E = \int \dot{E}_{\rm sd}dt$ from the magnetar wind as a function of $B_{\rm d}$ and $P_{0}$ for two values of the fall-back time, $t_{\rm fb} = 10$ s (left panel) and $t_{\rm fb} = 10^{5}$ s (right panel), and adopting the \citet{Piro&Ott11} coupling.  These two cases approximately cover the parameter space relevant to magnetars powering long GRBs and SLSNe, respectively.  Shown for comparison with horizontal lines are the total rotational energy of non-accreting magnetars with the same initial spin period, $E_{\rm rot}(P_{0})$ (eq.~\ref{eq:Erot}).

For relatively weak magnetic fields $B_{\rm d}$ and slow rotation $P_0 \gtrsim $ few ms, spin-up from accretion increases the extracted rotational energy as compared to an otherwise equivalent non-accreting case.  This has the important implication that the range of magnetar properties capable of producing GRBs or SLSNe is greatly enhanced by the effects of accretion.  On the other hand, for large $B_{\rm d}$ and small $P_0$, accretion can act to decrease the effective rotational energy reservoir of the magnetar wind via the propeller mechanism.  Across most of the parameter space, $E$ does not greatly exceed the maximum initial rotational period $E_{\rm rot}(P_{0}) \approx $ few $\times 10^{52}$ ergs for an isolated magnetar rotating near break-up.

We further illustrate the latter point by analytically estimating the maximum rotational energy which can be extracted from a magnetar under the constraint that the total accreted mass be $M_{\rm acc} \lesssim 0.8M_{\odot}$.  For $\dot{M} \lesssim \dot{M}_{\rm c}$ accretion either removes angular momentum from the magnetar or has no effect on its spin-down evolution as compared to the non-accreting case.  For $\dot{M} \gtrsim \dot{M}_{\rm c}$ the magnetar gains angular momentum and quickly approaches the equilibrium state $\dot{M} \simeq \dot{M}_{\rm c}$ ($P = P_{\rm eq}$) for which the wind luminosity obeys $\dot{E}_{\rm sd} = \dot{E}_{\rm sd,eq}$ (eq.~\ref{eq:EdotPeq}).  Substituting $\dot{M} = 0.8M_{\odot}/\Delta t$, the extracted rotational energy in a time interval $\Delta t$ is
\be
 E = \dot{E}_{\rm sd,eq}\Delta t = 6.7\times 10^{52}\,B_{15}^{-6/7}(\Delta t/10{\rm s})^{-3/10}M_{1.4}^{12/7}{\rm ergs}.
\ee
The shortest allowed value of $(\Delta t)_{\rm min} = 9.3\,{\rm s}B_{15}^{-2}M_{1.4}^{1/2}$ (largest $E$) is set by the requirement that $\dot{M} = \dot{M}_{\rm ns}$, since for $\dot{M} \gtrsim \dot{M}_{\rm ns}$ the spin-down rate no longer increases with increasing $\dot{M}$.  Thus, we have a maximum extractable energy
\be
E_{\rm max} = \dot{E}_{\rm sd,eq}(\Delta t)_{\rm min} = 6.7\times 10^{52}\,B_{15}^{-0.26}M_{1.4}^{1.56}{\rm ergs},
\label{eq:Erotmax}
\ee
which is less than a few times higher than the maximum initial rotational energy.  Equation \ref{eq:Erotmax} should also be augmented to include the remaining rotational energy after the accretion subsides, which, though comparable to $E_{\rm max}$, will emerge over longer timescales due to the slower spin-down of a non-accreting magnetar.

\subsection{$^{56}$Ni Production in GRB Supernovae}
\label{sec:56Ni}

In order to explain the large masses $\gtrsim 0.2-0.3M_{\odot}$ of radioactive $^{56}$Ni needed to power the light curves of  GRB SNe (e.g.~\citealt{Mazzali+14}) by shock-heating the inner layers of the progenitor star, the central engine outflow must inject $E \sim 10^{52}$ ergs of energy on a short timescale of $\lesssim 1$ s following the explosion (e.g.~\citealt{Barnes+17}).  Without accretion-induced enhancement of the spin-down power, this would require a magnetar with a short initial spin period $P_{0} \sim 1$ ms and a very strong magnetic field $B_{\rm d} \gtrsim 10^{16}$ G in order to obtain a spin-down time $t_{\rm sd,3}$ of less than a few seconds \citep{Suwa&Tominaga15}.  However, such a magnetar would spin-down quickly, sapping the rotational energy needed to power the GRB jet at later times ($\S\ref{sec:GRB}$).  

On the other hand, when the accretion enhancement of the spin-down rate is taken into account, from equation (\ref{eq:edotsd}) we see that for $P \sim 1$ ms a magnetic field $B_{\rm d} \gtrsim 2\times 10^{15}$ G is sufficient for a jet power $\dot{E}_{\rm sd} \gtrsim 10^{52}$ ergs s$^{-1}$, provided the accretion rate is close to the monopole spin-down limit, $\dot{M} \gtrsim \dot{M}_{\rm ns} \approx 0.086B_{15}^{2}M_{1.4}^{-1/2} M_{\odot}$ s$^{-1}$.  Even at such a high accretion rate, the total mass accreted by the magnetar of $\sim 0.1M_{\odot}$ over the necessary timescale $\sim 1$ s would be insufficient to instigate its collapse to a black hole.

\begin{figure*}[!t]
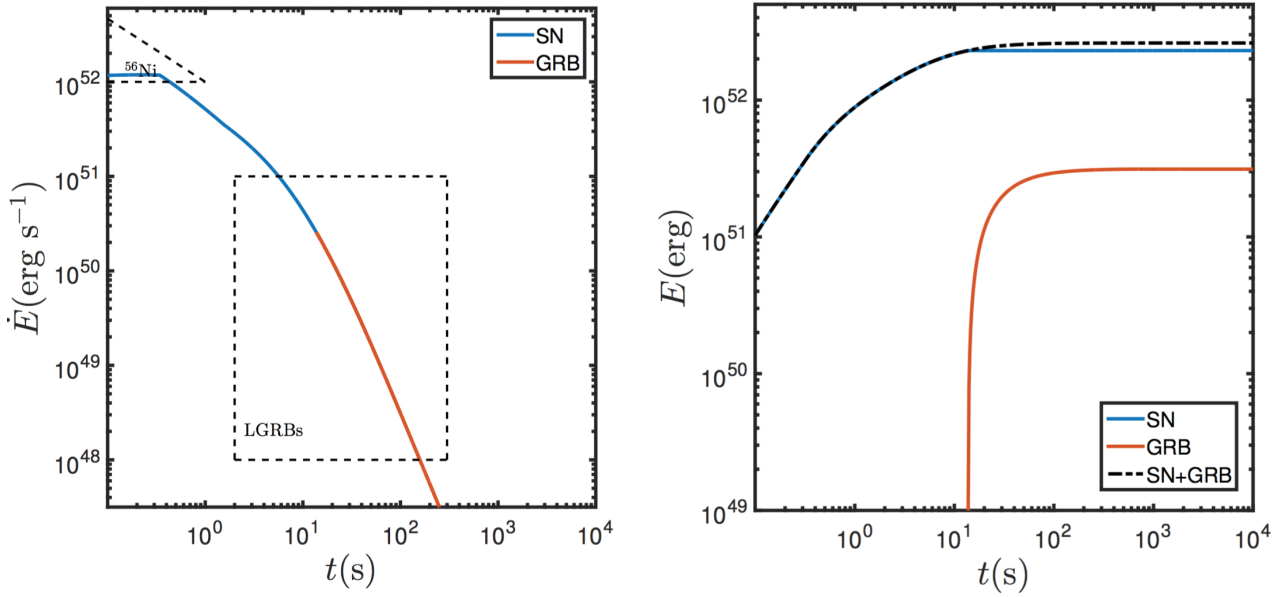

\includegraphics[width=0.48\textwidth]{GRBSNEdot.pdf}
\includegraphics[width=0.48\textwidth]{GRBSNE.pdf}
\caption{\footnotesize The magnetar wind can both power the supernova (by contributing to its kinetic energy and by shock-heating the inner layers of the star to produce sufficient $^{56}$Ni) as well as the subsequent GRB jet.  Here we show a magnetar with $B_{\rm d} = 2\times 10^{15}$ G and $P_{0} = 1$ ms that experiences a hypothetical two-component fall-back history comprised of separate accretion episodes with $t_{\rm fb,1} \approx 0.1$ s, $M_{\rm fb,1} \approx 0.5M_{\odot}$ and $t_{\rm fb,2} \approx 5$ s, $M_{\rm fb,2}\approx 0.3M_{\odot}$, respectively, where each component follows the temporal form given by equation (\ref{eq:mdotfb}).  Energy released when the magnetization of the magnetar outflow is in the range $100\leq \sigma \leq 3000$ contributes to the GRB jet (red curves; see \S\ref{sec:GRB}), while at earlier times when $\sigma \le 100$ the wind contributes to powering the SN (blue curves).  As shown in the left panel, the jet power is $\dot{E}_{\rm sd} \approx 10^{52}\,\mbox{ergs s}^{-1}$ in the first $\lesssim 0.5$ s, sufficient to produce the requisite large quantity of $^{56}$Ni by shock-heating the inner layers of the supernova ejecta (e.g.~\citealt{Suwa&Tominaga15,Barnes+17}), yet the jet power remains $\gtrsim 3\times 10^{50}\mbox{ergs s}^{-1}$ for a timescale $\sim 10$ s typical of the duration of a long GRB.  The right panel shows the cumulative energy released in the SN and GRB components, which total to $E_{\rm SN}=2\times 10^{52}$ ergs and $E_{\rm GRB}=3\times 10^{51}$ ergs, respectively. }
\label{fig:Ni56}
\end{figure*}

To illustrate this point, Fig.~\ref{fig:Ni56} considers a hypothetical mass accretion history characterized by two separate fall-back episodes.  The first episode has $t_{\rm fb} \approx 0.1$ s and carries a mass $M_{\rm fb} \approx 0.5M_{\odot}$, while the second component has $M_{\rm fb}\approx 0.3M_{\odot}$ and $t_{\rm fb} \approx 5$ s.  As shown in the left panel of Fig.~\ref{fig:Ni56}, the jet power reaches values of $\dot{E}_{\rm sd} \approx 10^{52}$ ergs s$^{-1}$ on timescales of $t \lesssim$ 0.5 s, while remaining $\gtrsim 3\times 10^{50}$ ergs s$^{-1}$ at $t \gtrsim 10$ s.  Such a jet evolutionary history can account for the large $^{56}$Ni masses of GRB-SNe, while also producing a GRB jet of sufficient power over longer timescales.  Assuming that the spin-down energy released when the magnetization of the outflow obeys $100\leq \sigma \leq 3000$ goes into powering the GRB jet (see \S\ref{sec:GRB}), while the remainder goes into feeding the kinetic energy of the SN explosion,  we find total energies of the SN and GRB components of $E_{\rm SN} \approx 2\times 10^{52}$ ergs and $E_{\rm GRB} \approx 3\times 10^{51}$ ergs, respectively; $90\%$ of the later is released within a time $\lesssim 70$ s, typical of a long GRB duration.

\subsection{GRB Jet Evolution}
\label{sec:GRB}
The wind from the magnetar is a promising source for feeding the relativistic GRB jet \citep{Thompson+04}, which escapes from the stellar ejecta on timescales of $\gtrsim 5-10$ s in the case of normal long GRBs (e.g.~\citealt{Bromberg+11,Aloy+18}).\footnote{The timescale for jet escape is shorter in the case of NS mergers due to the small ejecta mass (e.g.~\citealt{Bucciantini+12,Bromberg+17}) but longer in ultra-long GRBs due to the lower jet luminosity and/or more radially-extended progenitor star (\citealt{Quataert&Kasen12,Margalit+18}).}  Once a successful jet is established through the star, non-axisymmetric instabilities within the jet appear to have only a relatively weak impact on its structure \citep{Bromberg&Tchekhovskoy16} and a large fraction of the power carried by the magnetar wind will be placed into the luminosity of the jet, $L_{\rm j} = \dot{E}_{\rm sd}$ (\citealt{Bucciantini+09}).  

A key property of the jet is its magnetization, or maximum attainable bulk Lorentz factor, given by
\be
\sigma = \frac{L_{\rm j}}{\dot{M}_{\rm j} c^{2}},
\label{eq:sigma}
\ee
where $\dot{M}_{\rm j}$ is the rate at which baryon mass leaves the NS surface along the open magnetic flux (the red region in Fig.~\ref{fig:cartoon}).

For a weakly-magnetized NS, the neutrino wind is approximately spherical in geometry, and the mass loss rate due to neutrino ablation is approximately parameterized by \citep{Metzger+11}
\be
\dot{M}_{\nu,0}(t) = 3\times 10^{-4}(1 + t/t_{\rm kh})^{-5/2}e^{-t/t_{\rm thin}}\,M_{\odot}\,{\rm s^{-1}},
\label{eq:mdotnu}
\ee
where $t_{\rm kh} \approx 2$ s is the Kelvin-Helmholtz cooling timescale of the proto-NS and we have assumed a NS cooling evolution appropriate for a NS of mass $M_{\rm ns} = 2M_{\odot}$ (\citealt{Pons+99}; see \citealt{Metzger+11} for details). Absent the effects of accretion, the mass loss rate cuts off abruptly after the NS becomes optically thin to neutrinos, on a timescale of $t_{\rm thin} \approx 10-30$ s; hereafter we take $t_{\rm thin} = 20$ s.

Accretion introduces an  additional source of neutrino-driven mass loss from the neutrinos produced as matter settles onto the NS from the magnetic accretion column \citep{Piro&Ott11}.  This additional neutrino luminosity $L_{\nu,\rm acc}\approx GM_{\rm ns}\dot{M}/R_{\rm ns}$ will irradiate the polar cap (e.g.~\citealt{Zhang&Dai10}), where the outflow is being driven outwards along the open field lines (Fig.~\ref{fig:cartoon}).  This contributes an additional mass loss rate of
	\be 
	\dot{M}_{\nu, \rm acc}=1.2\times 10^{-5}  M_{1.4} \dot{M}_{-2}^{5/3} M_{\odot} \mbox{s}^{-1},
	\ee
which decays on the fall-back time-scale $t_{\rm fb}$ instead of $t_{\rm thin}$; thus $\dot{M}_{\nu, \rm acc}$ always exceeds $\dot{M}_{\nu,0}$ at sufficiently late times.

Above the magnetar surface, magnetic pressure greatly exceeds the thermal pressure from neutrino heating (e.g.~\citealt{Thompson+04,Vlasov+14}),  Thus, only the fraction $f_{\Phi}$ of the magnetar surface threaded by open magnetic flux is open to mass outflow (however, see \citealt{Thompson03,Thompson&UdDoula17}).  The baryon loading is thus suppressed relative to the case of a weakly-magnetized wind by a factor
\begin{eqnarray}
&&\frac{\dot{M}_{\rm j}}{\dot{M}_{\nu}} = f_{\Phi}f_{\rm cent} \nonumber \\
 &\simeq& \left\{
\begin{array}{lr} 0.5f_{\rm cent}  & \dot{M} \gtrsim \dot{M}_{\rm ns} \\
\frac{R_{\rm ns}}{2R_{\rm m}} \approx 0.27f_{\rm cent} B_{15}^{-4/7}\dot{M}_{-2}^{2/7}M_{1.4}^{1/7}& \dot{M}_{\rm lc}  \lesssim \dot{M} \lesssim \dot{M}_{\rm ns}, \\
\frac{R_{\rm ns}}{2R_{\rm lc}} \approx 0.13f_{\rm cent}P_{\rm ms}^{-1} & \dot{M} \lesssim \dot{M}_{\rm lc}, \\
\end{array}
\label{eq:mdotratio}
\right. ,
\end{eqnarray}
where $\dot{M}_{\nu} = \dot{M}_{\nu,0} + \dot{M}_{\rm \nu, acc}$ and we have taken $f_{\Phi} = 0.5$ for $R_{\rm m} = R_{\rm ns}$ to account for the fact that the geometrically-thick accretion disk covers roughly half of the NS surface in this limit.  Here
\be f_{\rm cent} = \exp[(P_{\rm c}/P)^{1.5}];\,\,\,\,\,P_{\rm c} \simeq 2.7\,\sin\theta_{\rm lc} M_{1.4}^{-1/2}\,{\rm ms}
\ee
accounts for the enhancement to the mass-loss rate due to centrifugal forces in proto-NS atmosphere \citep{Metzger+07}, where again $\theta_{\rm lc} =$sin$^{-1}(R_{\rm ns}/R_{\rm m})^{1/2}$ is the polar latitude of the last closed field line; the distance measured from the rotation axis to this location on the NS surface defines the centrifugal ``lever arm" at the base of the wind where most of the mass flux emerges.

\begin{figure*}[!t]
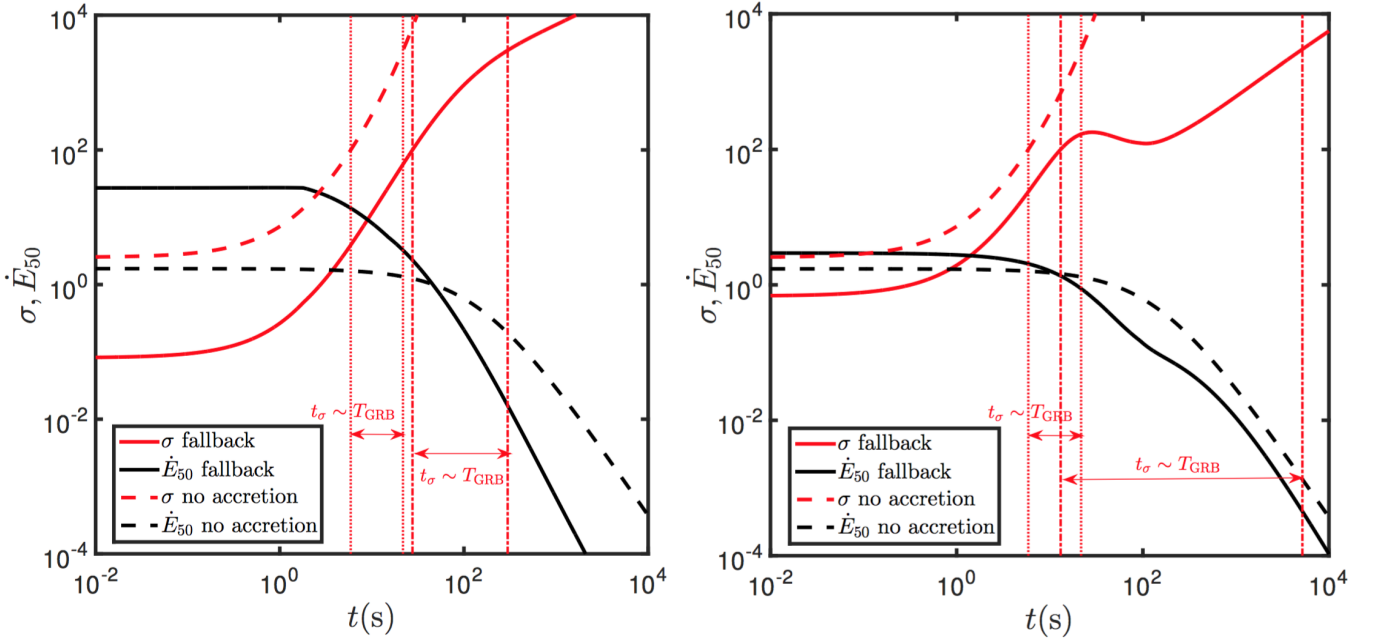

\includegraphics[width=0.5\textwidth]{Edotsigma.pdf}
\hspace{0.0cm}
\includegraphics[width=0.5\textwidth]{Edotsigma2.pdf}
\vspace{-0.4cm}
%\caption{\footnotesize 10}
%\includegraphics[width=0.5\textwidth]{misc1000.pdf}
%\hspace{0.0cm}
%\includegraphics[width=0.5\textwidth]{radii1000.pdf}
%\vspace{-0.4cm}
\caption{\footnotesize The additional baryon-loading from fall-back accretion can appreciably lengthen the duration of the prompt GRB emission.  This figure shows the jet power $L_{\rm j}(t) = \dot{E}_{\rm sd}$ (in units of $10^{50} \mbox{ergs s}^{-1}$; black curve) and magnetization $\sigma(t)$ (red curve) for a magnetar with $B_{\rm d} = 10^{15}$ G and an initial spin of $P_0=1$ ms, accreting 0.8$M_{\odot}$ in fall-back matter over a timescale $t_{\rm fb} = 3$ s (top panel) and $t_{\rm fb} = 300$ s (bottom panel).  Dashed lines show the same model, but neglecting the effects of fall-back.  Vertical lines indicate the duration of the GRB prompt emission, defined as the interval $t_\sigma \sim T_{\rm GRB}$ over which $100\leq \sigma\leq 3000$.  The effect of fall-back accretion is to length the GRB duration from $\lesssim 100$ s for an isolated magnetar to $\sim 300-5000$ s in the accretion case.}
\label{fig:Edotsig}
\end{figure*}

Figure \ref{fig:Edotsig} shows the evolution of the jet power and magnetization as a function of time for a magnetar with $B_{\rm d} = 10^{15}$ G, $P_{0} = 1$ ms which accretes 0.8$M_{\odot}$ in fall-back over a timescale $t_{\rm fb} = 3$ s (left panel) and $t_{\rm fb} = 300$ s (right panel).  In addition to the enhancement of the spin-down rate discussed earlier, the additional baryon-loading of the jet due to accretion causes the jet magnetization to rise much slower at late times as compared to the isolated (non-accreting) case.  

GRB prompt emission models typically require high but not excessive magnetization. Since $\sigma$ is an upper limit on the terminal Lorentz factor of the jet, $\Gamma$, compactness arguments impose a strong limit of $\sigma\geq \Gamma\gtrsim 100$ \citep{Fenimore1993,Woods1995,Lithwick&Sari01}. In many dissipation models $\sigma$ is in fact expected to be comparable to $\Gamma$. The latter is estimated from peaks in GRB afterglows and found to reside in the range $100\lesssim \Gamma \lesssim 3000$ \citep{Liang2010,Lu2012,Ghirlanda2012}. In models based on gradual magnetic reconnection, the same range of $\Gamma,\sigma$ is also required from a purely theoretical perspective \citep{Beniamini&Giannios17}.  

Because of this restricted `useful' range in $\sigma$, the more gradual rise in $\sigma$ from an accreting magnetar acts to significantly extend the duration of the prompt emission phase.  Without fall-back accretion, the timescale $t_{\sigma}$ over which the magnetization obeys $10^{2} \lesssim \sigma \lesssim 3000$ is limited to $t_{\sigma} \lesssim few \times t_{\rm thin} \lesssim$ tens of seconds.  However, when the effects of accretion are included, $t_{\sigma}$ can be significantly lengthened; for the example shown in the right panel of Fig.~\ref{fig:Edotsig}, $t_{\sigma}$ increases from $\approx 5$ s for $\dot{M} = 0$ to $\approx 5000$ s with accretion present.  

\begin{figure*}[!t]
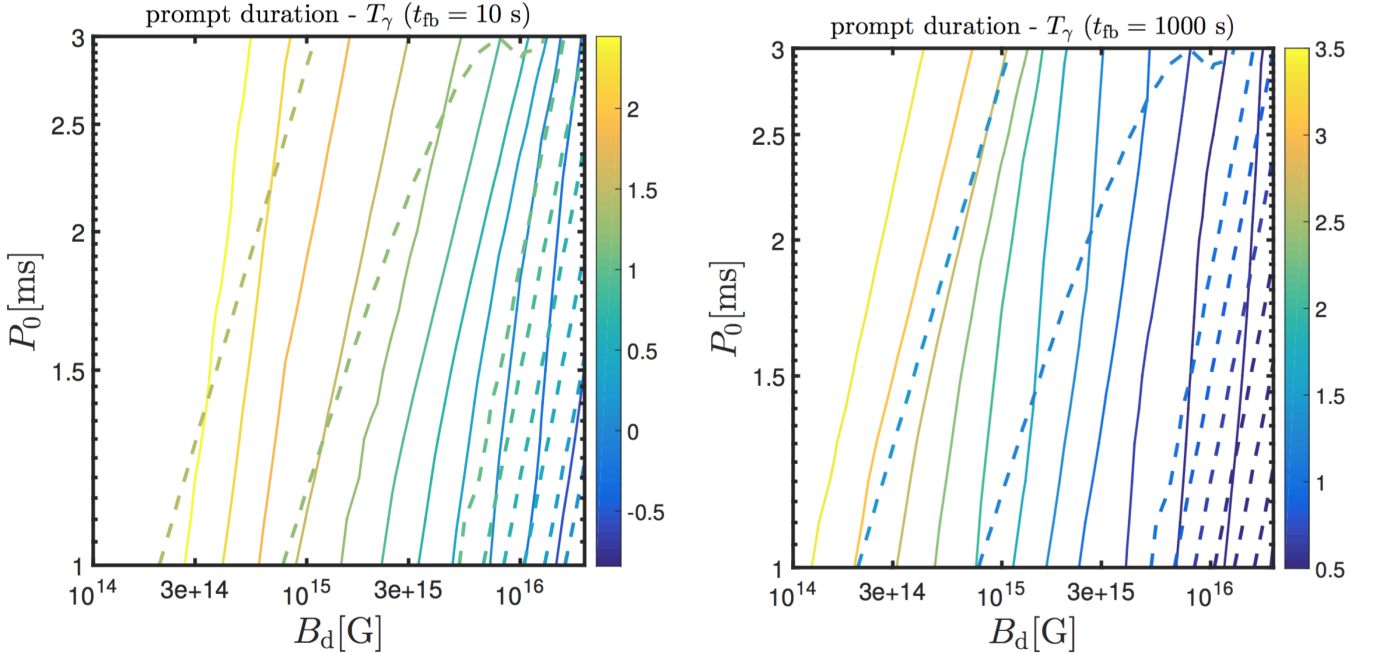

	\includegraphics[width=0.5\textwidth]{tGRB.pdf}
	\hspace{0.0cm}
	\includegraphics[width=0.5\textwidth]{tGRB2.pdf}
	\vspace{-0.4cm}
	\caption{\footnotesize Accretion can substantially the lengthen the duration of the prompt GRB emission across a wide range of the magnetar parameter space $B_{\rm d}-P_{0}$.  This figure shows the total duration of the GRB prompt emission, $T_{\rm GRB}$, which we take to be the maximum between $t_{\sigma}$ (the time interval over which the jet magnetization is in the range $\sigma=100-3000$ necessary to produce prompt emission) and the magnetar spin-down time $t_{\rm sd}$.  We assume an initial NS mass of $1.4M_{\odot}$, an accreted mass of $0.8M_{\odot}$ and mass fall-back times $t_{\rm fb} =10$ s (top panel) or $t_{\rm fb}=1000$ s (bottom panel). Dashed lines depict the equivalent contours for a case with zero accretion. Note the change in the color bar range between the top and bottom figures.}
	\label{fig:tGRB}
\end{figure*}

Figure \ref{fig:tGRB} shows $T_{\rm GRB}=\min(t_{\sigma},t_{\rm sd})$, a proxy for the GRB duration,  in the parameter space of $B_{\rm d}$ and $P_{0}$, taking into account the effects of fallback accretion.  This definition accounts for the fact that $t_{\sigma}$ only controls the duration of the GRB prompt emission when it does not exceed the spin-down timescale of the magnetar, $t_{\rm sd}$.  Shown for comparison with dashed lines are the predicted GRB durations without accretion.  When the effects of accretion are included, the GRB duration for $B_{\rm d} \sim 10^{14}-10^{15}$ G is increased by $\gtrsim 1-2$ orders of magnitude as compared to the non-accreting case.  This finding has the implication that the magnetar model can accommodate ultra-long GRBs with prompt emission that lasts for a duration $T_{\rm GRB} \sim 10^{3}-10^{4}$ s (e.g.~\citealt{Greiner+14,Levan15}).  Since the spectral energy distribution of the GRB depends in most models on the jet magnetization, accretion acts to reduce the predicted rate of spectral evolution during the burst.  This places the magnetar model in more comfortable agreement with time-resolved observations of GRB spectra, which generally show at most moderate hard-to-soft evolution (e.g.~\citealt{Kargatis+94}).

\begin{figure}[!t]
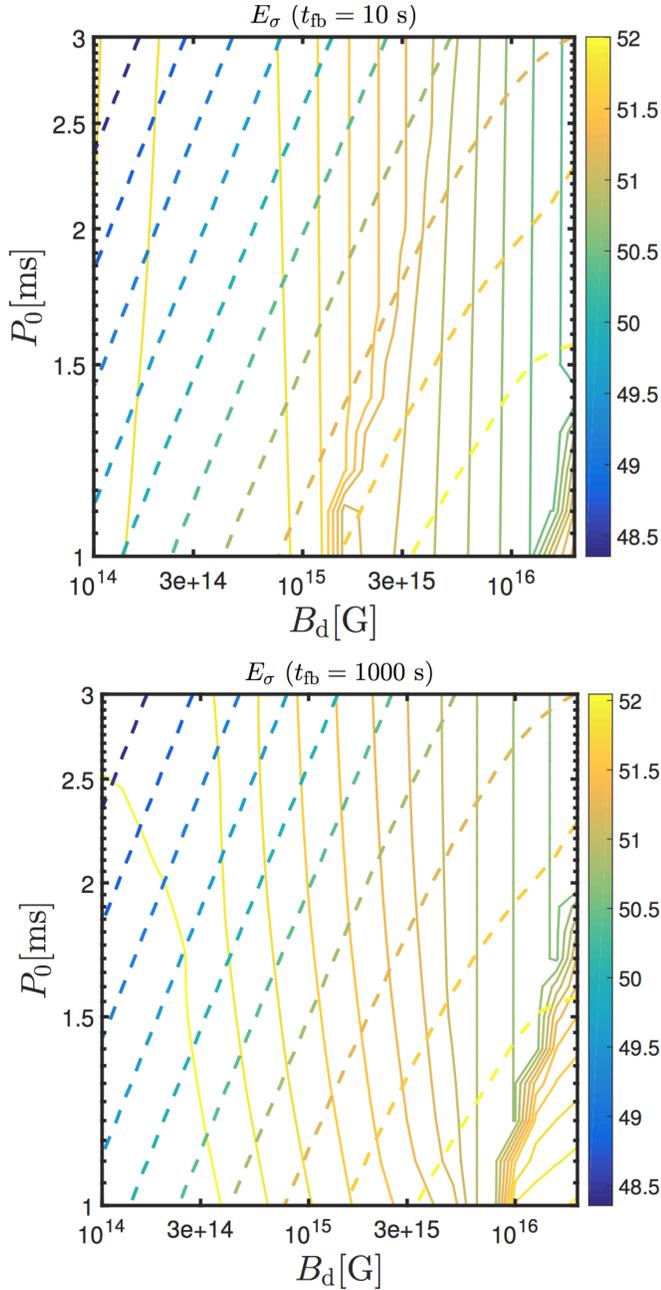

	\includegraphics[width=0.5\textwidth]{Esigma.pdf}
	\hspace{0.0cm}
	\includegraphics[width=0.5\textwidth]{Esigma2.pdf}
	\vspace{-0.4cm}
	\caption{\footnotesize Accretion can enhance the total energy released during the prompt GRB emission across a wide range of the magnetar parameter space $B_{\rm d}-P_{0}$.  This figure shows the total rotational energy released by the magnetar wind in the magnetization range $\sigma=100-3000$ necessary to produce prompt emission (e.g.~\citealt{Beniamini&Giannios17}). We assume an initial NS mass of $1.4M_{\odot}$, an accreted mass of $0.8M_{\odot}$ and $t_{\rm fb} =10$ s (top panel) or $t_{\rm fb}=1000$ s (bottom panel). Dashed lines depict the equivalent contours for a case with no accretion.}
	\label{fig:Esig}
\end{figure}

An upper limit on the prompt GRB energy is given by $E_{\sigma}$, the total rotational energy extracted from the magnetar as the wind evolves from $\sigma=100$ to $\sigma=3000$ \citep{Beniamini+17}, as shown in Figure \ref{fig:Esig}.  As in the case of the total extracted rotational energy, we find that $E_{\sigma} \lesssim 10^{52}$ergs across the parameter space of surface magnetic field and initial spin period.  Thus, even when including the effects of fall-back accretion, the total available energy for the GRB phase is limited to values similar to the isolated magnetar case.

\subsection{GRB with Long Precursor Gaps}
\label{sec:precursors}
While the magnetization of the jet powered by an isolated, non-accreting magnetar rises monotonically (e.g.~\citealt{Metzger+11}), the evolution of $\sigma(t)$ when the effects of accretion are included (\S \ref{sec:GRB}) can be substantially more complex and even non-monotonic, depending on the relative ordering of the timescales $t_{\rm fb},t_{\rm kh},t_{\rm thin},t_{\rm eq}$.  For $B_{\rm d} \gtrsim 7\times 10^{15}$G, the value of $\sigma$ reaches high values at early times $\geq 100$, before decreasing to values $\sigma \approx 30$ over the next few seconds as the magnetar quickly spins down due to accretion, before eventually rising up again to large values $\sigma \gtrsim 100$ as the magnetar cools and the neutrino-driven baryon loading of the jet subsides.  If the jet magnetization must be in the critical range $100\leq \sigma \leq 3000$ in order to produce prompt gamma-ray emission, then this complex $\sigma$ evolution would imprint an equally complex GRB light curve.  One would expect ``precursor" emission during the brief early high-$\sigma$ phase, followed by an extended gap without prompt emission when $\sigma$ is low, which is then finally accompanied by a second GRB emission episode as $\sigma$ again rises at late times. 

\begin{figure*}[!t]
\includegraphics[width=1\textwidth]{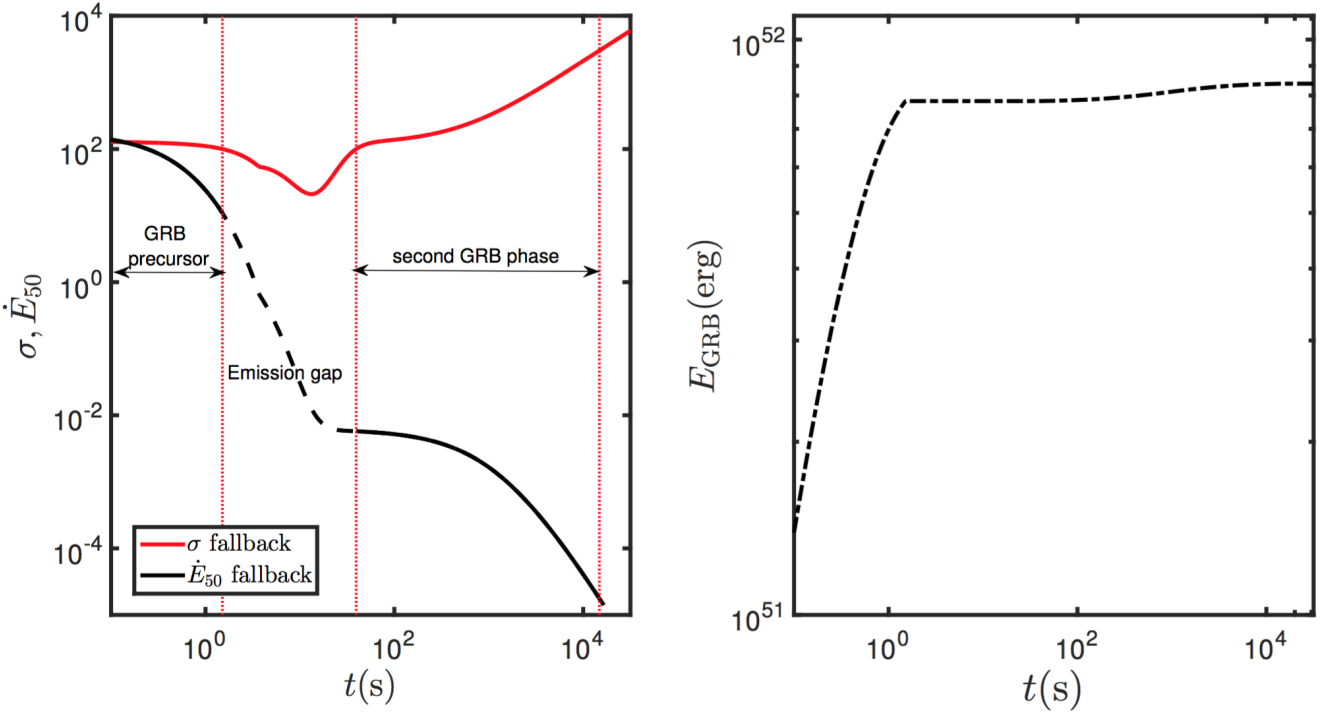}
\caption{\footnotesize Magnetar jets can produce precursor episodes, followed by large gaps before the ``main" emission episode.  This figure shows the evolution of the spin-down power $\dot{E}$ and magnetization $\sigma$ of a magnetar wind for $B_{\rm d} =10^{16}$G, $P_0=1$ ms which accretes $0.8M_{\odot}$ over a time $t_{\rm fb}=1000$.  Energy released with a magnetization of $100\leq \sigma \leq 3000$ is assumed to contribute to the GRB.  Note that the non-monotonic evolution of $\sigma$ results in the creation of a long temporal gap between the precursor and main emission episode.  The right panel shows the cumulative energy deposited in the GRB phase $E_{\rm GRB}$.}
\label{fig:gapGRB}
\end{figure*}

Figure \ref{fig:gapGRB} provides an explicit example of this behavior for a magnetar with $B_{\rm d} =10^{16}$ G and $P_0=1$ ms that accretes $0.8M_{\odot}$ over a time $t_{\rm fb} =1000$ s.  In this case the precursor phase lasts for $1.5$ s and releases $7.8\times 10^{51}$ ergs.  This is then followed by a $\approx 40$ s gap, after which a second GRB phase commences, which releases an additional $6\times 10^{50}$ ergs over a duration of $\sim 10^4$ s.  Although future work is needed to determine whether such a scenario can reproduce observed GRB precursors in detail, the jet evolution of an accreting magnetar is clearly substantially more complex than from an isolated one, helping to explain the wide diversity in GRB light curves.

\subsection{Binary NS Mergers}
\label{sec:BNS}

The merger of two NSs in a binary can result in the formation of a massive NS remnant, which is at least temporarily stable to gravitational collapse due to its rapid rotation \citep{Metzger+08,Bucciantini+12,Rowlinson+13,Giacomazzo&Perna13,Zhang13,Gao+15}.  As a result of the large orbital angular momentum of the initial binary, the remnant is rotating at close to its centrifugal break-up limit of $P_{0} \lesssim 1$ ms and will possess a strong magnetic field, at least on small spatial scales (e.g.~\citealt{Price&Rosswog06}).  

As in the core collapse case, such a magnetar remnant will accrete matter over an extended period after the merger, as marginally-bound high angular-momentum material falls back into a disk surrounding the merger and the remnant disk feeds matter onto the NS \citep{Rosswog07,Metzger+09,Metzger+10b,Fernandez&Metzger13,Metzger&Fernandez14,Fernandez+17}.  The total accreted mass is typically in the range $M_{\rm acc} \sim 0.01-0.1M_{\odot}$ and and the peak accretion rate is $\sim 0.1-1M_{\odot}$ s$^{-1}$, where the characteristic ``fall-back" time $t_{\rm fb} \sim 0.1$ s may in this case be controlled by the viscous timescale of the torus.

For $P \simeq 0.8$ ms and $M = 2.4M_{\odot}$, we see that $\dot{M}_{\rm ns} \simeq 4\dot{M}_{\rm c} \approx 0.07B_{15}^{2}M_{\odot}$ s$^{-1}$.  Therefore, unless $B_{\rm d} \gg 10^{15}$ G, the accretion rate is sufficiently high at early times to push the Alfv\'{e}n radius down by the disk to the NS surface.  For $t_{\rm fb} = 0.1$ s, and magnetic field strength of $B_{\rm d}$ = $3\times 10^{14} - 3\times 10^{15}$ G, we find that if the magnetar remnant is indefinitely stable (i.e. does not collapse to a black hole during its spin-down), then the total energy extracted in the magnetar wind is $\approx 1-2\times 10^{52}$ ergs (adoping the \citealt{Piro&Ott11} prescription), somewhat lower than the rotational energy $8 \times 10^{52}$ ergs which would be available were the magnetar spinning-down in isolation.  Eventually, as $\dot{M}$ continues to decrease an equilibrium will be established with $R_{\rm m} \simeq R_{\rm c}$ (as in the core collapse case), such that the spin-down luminosity will approach the decay $\dot{E}_{\rm sd} \propto t^{-1.9}$ (eq.~\ref{eq:latedecay}); however, in practice the time required to achieve this equilibrium is long, $\sim 3\times 10^{3}-3\times 10^{6}$ s, depending on $B_{\rm d}$.  \citet{Gompertz+14} and \citet{Gibson+17} explore a related model for the propeller-driven outflows from accreting magnetars as an explanation for the temporally-extended X-ray emission observed after some short GRBs. 

Instead of an indefinitely stable NS that spins down completely, a binary NS merger is more likely to produce a quasi-stable supramassive NS remnant, which collapses to a black hole once its spin period increases above a critical period $P_{\rm collapse}$, the precise value of which depends on the mass of the remnant relative to the theoretically uncertain maximum mass of a non-rotating NS.  \citet{Margalit&Metzger17} used energetic constraints on GW170817 to disfavor the formation of a long-lived supermassive NS remnant in this event by comparing the rotational energy released by an (isolated) NS as its spins down from an initial state close to the mass-shedding limit to the point of collapse, to observations constraining the kinetic energy of the GRB jet and kilonova ejecta.  However, these constraints would be weakened if the surrounding accretion disk would substantially reduce the rotational energy released by the magnetar prior to its collapse.  

\begin{figure}[!t]
\centering
\includegraphics[width=0.5\textwidth]{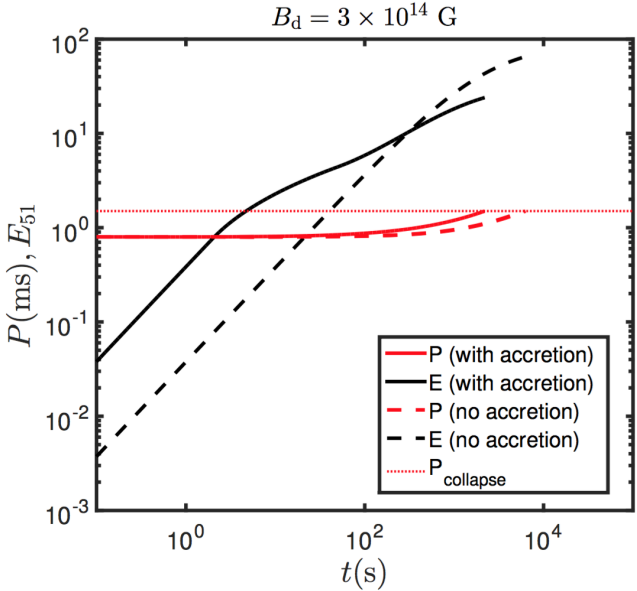}
\caption{\footnotesize Time evolution of the cumulative extracted wind energy (black curves) and spin period (red curves) of a supramassive NS formed from a NS-NS merger. We assume an initial NS mass and spin of $M_{\rm NS}=2.4M_{\odot},P=0.8$ms respectively as well as $M_{\rm fb}=0.1M_{\odot},t_{\rm fb}=0.1$s. The SMNS is assumed to collapse to a black hole once its period becomes slower than $P_{\rm collapse}=1.5$ ms (horizontal dashed line). }
\label{fig:NSmerger}
\end{figure}

Figure \ref{fig:NSmerger} shows the evolution of the cumulative extracted rotational energy for an accreting magnetar, which demonstrates that, due to the transfer of angular momentum from the star to the disk, the energy in the magnetized wind released prior to the point of collapse ($P = P_{\rm collapse}$) can be reduced by a factor of several relative to the otherwise identical case of an isolated magnetar.  However, while including the accretion effects could weaken the constraints of \citet{Margalit&Metzger17} slightly, it would not significantly alter their quantitative conclusions regarding the maximum mass of the NS. 

\subsection{SLSNe}
\label{sec:SLSNe}

Powering a SLSN through the birth of an isolated magnetar requires one with a lower magetic field strength, $B_{\rm d} \lesssim few\times 10^{14}$ G, for which the spin-down timescale is comparable to the photon diffusion timescale outwards through the expanding supernova ejecta \citep{Kasen&Bildsten10,Woosley10,Metzger+14}. The contribution of the magnetar to the peak luminosity of the supernova is approximately given by the spin-down power evaluated at the time when the ejecta first becomes transparent to visual radiation, i.e. the characteristic diffusion timescale \citep{Arnett82}
\be
t_{\rm peak} = \left(\frac{3\kappa M_{\rm ej}}{4\pi c v_{\rm ej}}\right)^{1/2} \approx 46\,{\rm d}\left(\frac{M_{\rm ej}}{10M_{\odot}}\right)^{1/2}\left(\frac{v_{\rm ej}}{10^{4}\,{\rm km\,s^{-1}}}\right)^{-1/2},
\ee
where here $\kappa \approx 0.1$ cm$^{2}$ g$^{-1}$, $M_{\rm ej}$ and $v_{\rm ej}$ are the opacity, total mass and mean velocity of the ejecta, respectively; the latter can be approximated from the ejecta kinetic energy
\be
\frac{M_{\rm ej}v_{\rm ej}^{2}}{2} = E_{\rm SN} + \int_{0}^{t_{\rm peak}}\dot{E}_{\rm sd},
\ee 
which in general is a combination of the initial explosion energy $E_{\rm SN} = 10^{51}$ ergs plus additional energy dumped into the ejecta by the magnetar prior to its transparency (\citealt{Metzger+15,Soker&Gilkis17}).  Thus, the peak luminosity can be estimated as
\be
L_{\rm peak} \simeq \dot{E}_{\rm sd}(t_{\rm peak}) + L_{\rm Ni}(t_{\rm peak}),
\label{eq:Lpeak}
\ee
where 
\be L_{\rm Ni} \approx 1.3\times 10^{43} {\rm  ergs \,s^{-1}} (M_{\rm Ni}/0.2M_{\odot})\exp(-t_{\rm peak}/8.8{\rm d})\ee is the minimal contribution to the luminosity arising from the radioactive decay of $^{56}$Ni of assumed mass $M_{\rm Ni}$.

\begin{figure}[!t]
\centering
\includegraphics[width=0.5\textwidth]{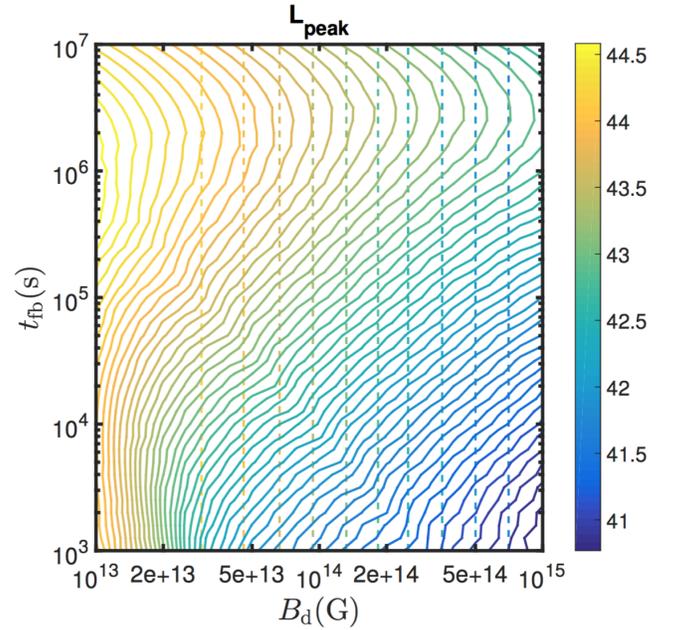}
\caption{\footnotesize Peak luminosity of a magnetar-powered SLSNe (eq.~\ref{eq:Lpeak}) including the effects of accretion, as a function of surface dipole field $B_{\rm d}$ and fall-back time of the debris $t_{\rm fb}$. All models assume an initial spin-period of $P_0 = 2$ ms, total accreted mass $M_{\rm acc} = 0.8M_{\odot}$, and supernova ejecta mass $M_{\rm ej} = 10M_{\odot}$.  Colored dashed lines depict the peak luminosity for the otherwise identical case with no accretion.}
\label{fig:SLSNe}
\end{figure}

Figure \ref{fig:SLSNe} shows the peak luminosity of magnetar-powered SLSNe as a function of $B_{\rm d}$ and $t_{\rm fb}$ for $P_{0} = 2$ ms, $M_{\rm fb} = 0.8M_{\odot}$, $M_{\rm ej} = 10M_{\odot}$. The peak luminosity is dominated by $\dot{E}_{\rm sd}(t_{\rm peak})$ and is typically in the range $L_{\rm peak} \sim 10^{42}-10^{45}\mbox{ergs s}^{-1}$.  Dashed lines show for comparison the luminosity in an otherwise equivalent case where accretion is negligible.  For magnetars with $t_{\rm fb} \ll t_{\rm peak} \sim 3\times 10^{6}$ s, the effects of accretion are to remove angular momentum from the magnetar and reduce the supernova luminosity, potentially below the SLSNe threshold $L_{\rm peak} \sim 10^{44}$ ergs s$^{-1}$.  On the other hand, for $t_{\rm fb} \gtrsim t_{\rm peak}$, the region of $B_{\rm d}-P_{0}$ parameter space capable of producing SLSNe can be expanded.  However, the stellar progenitors capable of producing such long fall-back times $t_{\rm fb} \gtrsim 10^{6}-10^{7}$ s are blue supergiants with extended hydrogen-rich outer envelopes \citep{Quataert&Kasen12}, in tension with the hydrogen-poor classification of many SLSNe.  Kinetic energy released in outflows from the accretion disk may provide a more promising source than the magnetar itself for powering SLSNe in the case of long fall-back times (e.g.~\citealt{Piro&Ott11,Dexter&Kasen13}).    

\section{Summary and Conclusions}
\label{sec:conclusions}

We have explored the effects of mass fall-back on the evolution of millisecond proto-magnetars formed in core collapse supernovae and neutron star mergers, comparing the evolution of their magnetized outflows and outflow-fed relativistic jets to the normally-considered case of an isolated (non-accreting) magnetar.  Broadly, we find that adding accretion to the picture substantially expands the possible behavior of magnetar engines, alleviating some (but not all) of the observational or theoretical drawbacks present in the isolated magnetar model.  

Our primarily conclusions are summarized as follows:
\begin{itemize}

\item Fall-back accretion can appreciably alter the spin-down evolution of magnetars, even when the accreted mass is insufficient to instigate the gravitational collapse of the magnetar to a black hole.  

Beyond enhancing (or reducing, in the propeller regime) the angular momentum of the star, the additional magnetic flux opened as the magnetosphere is compressed by accretion can greatly enhance the magnetar's spin-down luminosity (Fig.~\ref{fig:MdotP}).  One way to think about this is that accretion acts to increase the {\it effective} magnetic field strength entering the dipole spin-down formula by a factor (eqs.~\ref{eq:edotsd}, \ref{eq:edotsdcases})
\begin{eqnarray}
&&\frac{B_{\rm eff}}{B_{\rm d}} = \frac{R_{\rm lc}}{R_{\rm m}}  = \left(\frac{\dot{M}}{\dot{M}_{\rm lc}}\right)^{2/7} \nonumber \\
 &\simeq& 2.15 P_{\rm ms}B_{15}^{-4/7}\dot{M}_{-2}^{2/7}M_{1.4}^{-1/7}, \dot{M} \ge \dot{M}_{\rm lc}.
\label{eq:Beff}
\end{eqnarray}
The larger effective value of $B_{\rm d}$ when accretion is present has the implication that the magnetic field strength inferred by modeling the luminosity and timescale of the prompt emission phase (when $\dot{M}$ is high) may not match the values inferred for the same events at later times, e.g. from the X-ray plateau phase \citep{Rowlinson+13}, once accretion has subsided (Fig.~\ref{fig:Beff}).      

\begin{figure}[!t]
\centering
\includegraphics[width=0.5\textwidth]{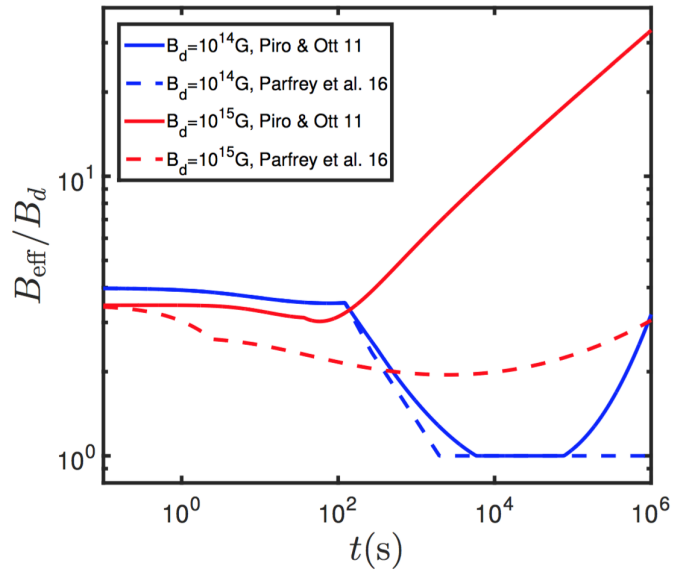}
\caption{\footnotesize Time evolution of the effective dipole magnetic field strength (eq.~\ref{eq:Beff}) for purposes of calculating the magnetar spin-down luminosity when the effects of accretion are included, as shown for different surface dipole field strengths and fall-back times (assuming a total accreted mass of $0.8M_{\odot}$).  A high value of $B_{\rm eff}$ at early times can explain a powerful GRB jet, which for the \citet{Parfrey+16} torque prescription, later decreases and could contribute to a long-lived X-ray plateau.}
\label{fig:Beff}
\end{figure}

\item{Accretion generally drives the spin period towards a value $P_{\rm eq}(t)$ corresponding to the condition $R_{\rm m} = R_{\rm c}$ (eq.~\ref{eq:Peq}).  However, the details of whether this quasi-equilibrium is achieved depends on the precise coupling between the accretion disk and the NS magnetosphere, which remains an area of active debate even in the comparatively well-studied case of accreting pulsars.  

When $P_{\rm eq}$ is obtained, the spin-down luminosity approaches a late-time decay $\dot{E}_{\rm sd} \propto t^{-\alpha}$, where $\alpha = 1.90-2.38$ (eq.~\ref{eq:latedecay}), moderately shallower or steeper than the standard $\dot{E}_{\rm sd} \propto t^{-2}$ prediction for an isolated magnetar.  Such differences might be observable on the post-maximum light curve decay of SLSNe \citep{Nicholl+16}, though this effect is degenerate with a decreasing fraction of the magnetar's energy being thermalized behind the ejecta (due to the growing fraction of UV, X-ray or gamma-ray radiation which directly escapes from the nebula without being reprocessed into optical emission; \citealt{Metzger+14}).  The power-law decay might also be imprinted on the late-time X-ray emission from GRBs (e.g.~\citealt{Rowlinson+13,Lasky+17}), though this association depends on the possibly dubious assumption that the spin-down power faithfully tracks the isotropic X-ray luminosity.  }

\item Accretion does not substantially enhance the maximum extractable rotational energy from a magnetar (Fig.~\ref{fig:totalenergy}) as compared to the case of an isolated (non-accreting) magnetar which is born maximally spinning, near break-up, $P_0 \simeq 1$ ms.  This is because the amount of accreted mass required to maintain $P = P_{\rm eq} \lesssim 1$ ms for many spin-down times exceeds that which would instigate collapse to a black hole (see eq.~\ref{eq:Erotmax} and surrounding discussion).  The concordance between the supernova energy scale of $\approx few \times 10^{52}$ ergs used by some authors to favor a magnetar origin for long GRB \citep{Mazzali+14} remains preserved, if not strengthened, when accretion is included.  

The propeller wind itself could carry away a substantial amount of kinetic energy (e.g.~\citealt{Piro&Ott11,Gompertz+14}).  However, such an outflow would be slower and more heavily baryon-loaded than the magnetar wind and thus cannot contribute to the ultra-relativistic GRB jet (though it could contribute to powering SLSNe for long fall-back times).  Tension thus remains between the magnetar model and the most energetic GRBs \citep{Cenko+11,Beniamini+17}.  

Perhaps more importantly, by contributing angular momentum to an initially slowly-spinning star, and by enhancing the spin-down luminosity compared to the isolated dipole (eq.~\ref{eq:tsd}, \ref{eq:Beff}), accretion does substantially increase the parameter space of magnetar birth properties ($B_{\rm d}, P_{0}$) which could give rise to the production of a long GRB jet  (Fig.~\ref{fig:Esig}).  Accretion thus reduces the gap between the properties of magnetars which were previously thought capable of giving rise to GRBs versus SLSNe \citep{Metzger+15,Margalit+18}.  The difference between GRB- and SLSN-producing magnetars could have as much or more to do with the fall-back accretion history they experience than with intrinsic differences in the magnetar birth properties.

\item The gravitational energy liberated by accretion results in a persistent source of neutrino irradiation of the magnetar polar cap, supplying an additional source of neutrino-driven mass-loss than would be present from an isolated cooling magnetar (Fig.~\ref{fig:cartoon}).  Accretion thus maintains the baryon loading of the magnetar jet in the requisite range (magnetization $\sigma \lesssim 10^{3}$) to produce prompt emission for timescales $\sim 10^{2}-10^{4}$ s, much longer than the Kelvin-Helmholtz cooling timescale of the star, thereby increasing the maximum duration of GRB prompt emission to the range of ultra-long GRBs (Fig.~\ref{fig:Edotsig}, \ref{fig:tGRB}).  This supports the possibility that GRB 111209A/SN2011kl \citep{Greiner+15} represents the transitional case of a magnetar with an intermediate spin-down timescale $t_{\rm sd} \sim 10^{4}$ s, which powered both the prompt GRB phase and later the SLSNe emission by re-energizing the supernova ejecta \citep{Metzger+15,Gompertz&Fruchter17}.  By slowing the rate at which the jet magnetization $\sigma(t)$ rises at late times, this also alleviates tension between the rapid spectral evolution of prompt GRB emission predicted by isolated magnetar models and observations \citep{Metzger+11,Beniamini+17}.

\item Accretion allows for more complex time evolution of the magnetar spin-down power and wind baryon-loading (magnetization), which can in turn enable more complicated GRB light curve behavior than the smooth monotonic evolution predicted for an isolated magnetar.  This additional freedom might help explain long GRBs with bright precursors which are followed by large temporal gaps before the main prompt emission episode (Fig.~\ref{fig:gapGRB}).  The enhanced spin-down rate of an accreting magnetar could also enable a large jet power in the first $\lesssim 1$ s after the supernova explosion (as needed to produce sufficient radioactive $^{56}$Ni through shock-heating of the progenitor envelope) while spinning down at a more gradual rate over the subsequent tens of seconds or longer as the accretion rate subsides, as needed for producing the longer duration of the GRB (Fig.~\ref{fig:Ni56}).

\item An accreting millisecond magnetar could also be present in the aftermath of a binary NS merger.  The accretion rate can be sufficiently high at early times to push the Alfven radius down to the neutron star surface and open the magnetosphere into a monopole-like field structure \citep{Metzger+18}.  However, because the disk is short-lived, its overall impact on the angular momentum and rotational energy budget of the magnetar is relatively minor.  Still, angular momentum extracted by the disk would reduce the energy released into the surrounding environment from the magnetar wind (Fig.~\ref{fig:NSmerger}).  This could moderately weaken energetic constraints which case be placed on the equation of state of the NS based on the type of compact remnant created in gravitational wave-detected mergers (\citealt{Margalit&Metzger17}).  It would also impact inferences about the existence of stable magnetar remnants following short gamma-ray bursts based on their late-time radio emission \citep{Metzger&Bower14,Horesh+16,Fong+16}.

\item The effect of mass accretion in magnetar-powered SLSNe is generally to reduce the rotational energy available to power the supernova if the fall-back time is shorter than the peak timescale of the SN of typically $\approx$ weeks.  Longer fall-back times could enhance the SLSNe emission; however, the more radially extended progenitor stars required to give such late fall-back should possess hydrogen envelopes, in tension with the Type I classification of engine-powered SLSNe (however, see \citealt{Sukhbold&Thompson17}).

\end{itemize}

\acknowledgements
BDM acknowledges support from NASA grants NNX16AB30G and NNX17AK43G issued through the Astrophysics Theory Program.  DG acknowledges support from NASA grants NNX16AB32G and NNX17AG21G issued through the Astrophysics Theory Program

%\bibliographystyle{yahapj}
%\bibliography{ms}

%\bibliographystyle{mn2e}
%\bibliography{ms}

\end{document}